\documentstyle[12pt,aaspp4,flushrt]{article}

\def\today{\ifcase\month\or January\or February\or March\or April\or May\or
  June\or July\or August\or September\or October\or November\or December\fi
  \space\number\day, \number\year}
\def\beq{\begin{equation}}
\def\eeq{\end{equation}}
\def\capt{\small \baselineskip 12pt }%
\def\bm{{\bf m}}
\def\bd{{\bf d}}
\def\br{{\bf r}}
\def\etal{{\it et al.\ }}

\def\eg{{\it e.g.}}
\def\ie{{\it i.e.}}

\def\kms{\ {\rm km\,s^{-1}}}
\def\hmpc{\,{\rm h^{-1}Mpc}}
\def\ihmpc{\, h\, {\rm Mpc^{-1}}}

\def\3hmpc{\, ( h^{-1} {\rm Mpc})^3}
\def\ln{{\rm ln}}
\def\log{{\rm log}}

\def\rlcdm{\Lambda{\rm CDM}\,} 
\def\rocdm{{\rm OCDM}\,} 
\def\lcdm{$\Lambda$CDM\,\,} 
\def\ocdm{OCDM\,\,} 

\def\h60{h_{60}}
\def\P{{\cal P}} 
\def\L{{\cal L}} 
\def\pmb#1{\setbox0=\hbox{#1}%
 \kern-.025em\copy0\kern-\wd0
 \kern.05em\copy0\kern-\wd0
 \kern-.025em\raise.0433em\box0}

\begin{document}
{\footnotesize \raggedleft FERMILAB-Pub-99/079-A \\}

\title{Large-Scale Power Spectrum and Cosmological Parameters\\
       from SFI Peculiar Velocities}

\author{Wolfram Freudling \altaffilmark{1,2}, Idit Zehavi \altaffilmark{3,4},\\
Luiz N. da Costa \altaffilmark{2,5}, Avishai Dekel \altaffilmark{3}, 
Amiram Eldar \altaffilmark{3}, Riccardo Giovanelli \altaffilmark{6}, \\
Martha P. Haynes \altaffilmark{6}, John J. Salzer \altaffilmark{7},  
Gary Wegner \altaffilmark{8}, 
and Saleem Zaroubi \altaffilmark{3} 
}

\altaffiltext{1}{Space Telescope-European Coordinating Facility, European
Southern Observatory, Karl-Schwarzschild Str. 2, D-85748 Garching b. 
M\"{u}nchen, Germany}
\altaffiltext{2}{European Southern Observatory, Karl-Schwarzschild Str. 2,
D-85748 Garching b. M\"{u}nchen, Germany}
\altaffiltext{3}{Racah Institute of Physics, The Hebrew University,
Jerusalem 91904, Israel} 
\altaffiltext{4}{NASA/Fermilab Astrophysics Group, Fermi National Accelerator
Laboratory, Box 500, Batavia, IL 60510-0500}
\altaffiltext{5}{Observatorio Nacional, Rua Gen. Jose Cristino 77,
 Rio de Janeiro, Brazil} 
\altaffiltext{6}{Center for Radiophysics and Space Research and National
Astronomy and Ionosphere Center, Cornell University, Ithaca, NY 14953}
\altaffiltext{7}{Dept. of Astronomy, Wesleyan University, Middletown, CT 06457}
\altaffiltext{8}{Dept. of Physics and Astronomy, Dartmouth college, Hanover, NH
03755}

\begin{abstract}

We estimate the power spectrum of {\it mass} density fluctuations from
peculiar velocities of galaxies by applying an improved
maximum-likelihood technique to the new all-sky SFI catalog.
Parametric models are used for the power spectrum and the errors, and
the free parameters are determined by assuming Gaussian velocity
fields and errors and maximizing the probability of the data given the
model.  It has been applied to generalized CDM models with and without
COBE normalization.  The method has been carefully tested using
artificial SFI catalogs.  The most likely distance errors are found to
be similar to the original error estimates in the SFI data.  The
general result that is not very sensitive to the prior model used is a
relatively high amplitude of the power spectrum.  For example, at
$k=0.1\ihmpc$ we find $P(k)
\Omega^{1.2} = (4.4\pm1.7)\times 10^3 \3hmpc$. An integral over the
power spectrum yields $\sigma_8 \Omega^{0.6} = 0.82 \pm 0.12$.
Model-dependent constraints on the cosmological parameters are
obtained for families of CDM models. For example, for COBE-normalized
\lcdm models (scalar fluctuations only), the maximum-likelihood result
can be approximated by $\Omega\, n^{2}\, {\h60}^{1.3} = 0.58 \pm
0.11$.  The formal random errors quoted correspond to the $90\%$
confidence level.  The total uncertainty, including systematic errors
associated with nonlinear effects, may be larger by a factor of $\sim
2$. These results are in agreement with an application of a
similar method to other data (Mark III).

\end{abstract}


\vfill
\eject

\section{INTRODUCTION}
\label{sec:intro}

In the standard picture of cosmology, structure originated from
small-amplitude density fluctuations that were amplified by
gravitational instability.  These initial fluctuations are assumed to
have a Gaussian probability distribution, fully characterized by their
power spectrum (PS).  On large scales, the fluctuations are expected
to be linear even at late times, still characterized by the initial
PS. Thus, the PS is a very useful statistics for large
scale-structure.  We focus on the PS rather than the correlation function 
(\eg, G\'orski \etal 1989) because the PS distinguishes more
clearly between the processes that affect structure formation on
different scales. It also has the advantage of being less sensitive to
assumptions regarding the mean density.

The PS has been estimated from several redshift surveys of galaxies
(see reviews by Strauss \& Willick 1995; Strauss 1998).  However, the
distribution of galaxies does not necessarily provide a direct
measurement of the underlying {\it mass} distribution; the PS
estimated from redshift surveys is contaminated by unknown ``galaxy
biasing".  Additional complications arise from redshift distortions,
triple-value zones and nonlinearities of the density field. This
also complicates a direct comparison of the correlation function
derived from galaxy density fluctuations with similar quantities derived 
from peculiar velocity measurements.  Therefore, it is
advantageous to estimate the mass PS directly from purely
dynamical data.  Another advantage of velocity over density data
is that they probe the density field on scales larger than the sample
itself, and that they are subject to weaker nonlinear effects.
It is therefore easier to obtain an approximation for the initial 
PS from the current velocity PS than from the current density PS. 

Direct estimation of the PS from reconstructed velocity or density
fields is complicated by the need to correct for the effects of large
noise, smoothing, and finite and nonuniform sampling (\eg, Kolatt \&
Dekel 1997). On the other hand, the likelihood analysis of peculiar
velocities, such as the one applied here, provides an appealing method
for estimating the mass PS since it is a straightforward statistic
acting on the `raw' data, without the need for processing such as
binning, smoothing, or applying a full POTENT reconstruction (Dekel,
Bertschinger \& Faber 1990).  It takes into account the measurement
errors and finite discrete sampling, and it utilizes much of the
information content of the data.  The simplifying assumptions made in
our main analysis are that the peculiar velocities are drawn from a
Gaussian random field, that the velocity correlations can be derived
from the density PS using linear theory, and that the errors in the
measurements are Gaussian. Other limitations of the
method are the need to assume some parametric functional form for the
PS, with a possible sensitivity of the results to the choice of this
model, and the fact that the likelihood analysis provides only
relative likelihood of the different models, not an absolute
goodness-of-fit.

PS estimates using likelihood analysis (Zaroubi \etal 1997) has been
obtained from the ``Mark III'' catalog of peculiar velocities (Willick
\etal 1997a), yielding relatively high values for the PS, in agreement
with the direct estimates from the ``POTENT'' reconstruction (Kolatt
\& Dekel 1997).  This result is still associated with large
uncertainties because the sampling of the data is sparse and
nonuniform, because the merging of data from several sources is
nontrivial, and because the distance errors in peculiar-velocity data
are relatively large.  Furthermore, the uncertainty in the assumed
distance errors always propagates into an uncertainty in the resultant
PS because the errors add in quadrature to the PS. It is therefore
important to analyze new data of certain improved qualities and to pay
special attention to the error estimates.

The data analyzed in the present paper are based on the new SFI
catalog of peculiar velocities of galaxies (Haynes \etal 1999, Wegner
\etal 1999), containing about 1300 field spiral galaxies with
Tully-Fisher  (TF) distances.  Most of the measurements in the SFI catalog
are new.  Data taken from the literature which are included in the
catalog, mostly those by Mathewson, Ford \& Buchhorn (1992), have been
recalibrated to match the new observations both for magnitude and line
width scale.  This procedure should minimize the effects of combining
different datasets, effects of significant concern in Mark~III (e.g.,
Willick \& Strauss 1998). The SFI catalog, though sparser than Mark
III in certain places, covers more uniformly the volume out to
$70\hmpc$.

The distances in the SFI catalog have been estimated using a linear TF
relation derived from a matching cluster sample (Giovanelli \etal
1997a, 1997b; SCI).  Possible deviations from the standard, linear TF
relation were ignored since no clear evidence for such deviations was
detected in the data. In addition, the sensitivity to such an effect
is small because of the the selection criteria of the SFI catalog.
  
The crucial issue of error estimate is addressed in two ways. First,
the fact that the SFI field sample is matched by the SCI cluster
sample of similar size allows a careful investigation of the
observational and internal scatter of the TF distances which provides
a good a priori estimate of the errors. These errors are adjusted for
an assumed difference in the scatter between field and cluster
galaxies. An additional adjustment of the scatter is due to our
bias-correction procedure.

A second and independent approach to estimate the errors is to include
them as an extra parameter in the likelihood analysis so that it also
determines the maximum-likelihood values for the errors. In that
approach, we use a parametric model for the errors which builds upon
the original estimates of width-dependent errors.

We address here the mass-density power spectrum as derived from
peculiar-velocity data, with or without Cosmic Microwave Background
(CMB) fluctuation data, but independent of the distribution of
galaxies in redshift space. We thus determine the quantity $P(k)
\Omega^{1.2}$ (where $P(k)$ is the density power spectrum and $\Omega$
is the cosmological density parameter), while we are free of
assumptions regarding the ``biasing" relation between galaxies and
mass. We can therefore measure a purely dynamical parameter such as
$\tilde\sigma_8\equiv\sigma_8 \Omega^{0.6}$ (where $\sigma_8$ is the
rms mass-density fluctuation in top-hat spheres of radius
$8\hmpc$). When assuming a priori a parametric functional form for the
mass PS, \eg, based on a generic CDM model, we can in fact determine a
combination of dynamical parameters such as $\Omega$ and the power
index $n$.

Investigations involving galaxy redshift surveys commonly measure a
different parameter that does involve galaxy biasing,
$\beta\equiv\Omega^{0.6}/b$ (where $b$ is the biasing parameter). The
parameters $\tilde\sigma_8$ and $\beta$ (at $8\hmpc$) are related via
$\sigma_{8\rm g}$, referring to the rms fluctuation in the galaxy
number density.  A number of measurements of $\beta$ have been carried
out so far, either based on redshift distortions of the IRAS 1.2 Jy
redshift survey (Fisher \etal 1995) or based on comparisons of this
redshift survey and the peculiar-velocity data. Most recent
velocity-velocity comparisons found values for $\beta$ in the range of
$0.5-0.7$ (Davis, Nusser \& Willick 1996; Willick \etal 1997b; da
Costa \etal 1998; Kashlinsky 1998; Willick \& Strauss 1998), while
density-density comparisons have lead to values as high as 0.9
(e.g. Sigad \etal 1998).  A determination of $\tilde\sigma_8$ from the
SFI data may help to clarify the situation.

In \S~\ref{sec:data} we describe the data and our method for
correcting Malmquist bias.  In \S~\ref{sec:method} we present the
method of analysis and the parametric models used as priors.  The
method is tested using mock catalogs in \S~\ref{sec:mock}.  The
estimated power spectra and the constraints on the cosmological
parameters are presented in \S~\ref{sec:results}.  The robustness of
the results is addressed in \S~\ref{sec:robust}.  We  
discuss our results and conclude in \S~\ref{sec:concl}.

\section{DATA}
\label{sec:data}

\subsection{Sample and Distance Errors}
\label{subsec:derrors}

The SFI sample is based on a wide-angle survey of Sbc-Sc galaxies with
I-band TF distances, covering declinations $\delta \ge -45^{o}$ and
galactic latitudes $b \ge 10^{o}$. The galaxy selection criteria
depend on redshift in order to ensure dense sampling at large
distances; the catalog consists of three zones of different diameter
limits and redshift limits. This data set was complemented south of
$\delta=-45^{o}$ with galaxies drawn from the Mathewson \etal (1992)
survey, carefully converted to the same system of magnitude and
line-width, and with the same set of corrections and selection
criteria applied to the whole sample (Giovanelli \etal 1997a,
1997b). The combined sample comprises of about 1300 field galaxies,
extending out to $7500 \kms$ in redshift, and quite isotropically
covering the whole sky except of the Galactic zone of avoidance.

Accurate estimation of the uncertainty $\Delta$ in the distance are
important both for the bias correction (see \S~\ref{subsec:bias}) and
for determination of the PS (see \S~\ref{sec:method}).  The
uncertainties are derived from the estimate of the scatter in the
observed TF relation.  We take advantage of the fact that the SFI
sample is matched by a similar cluster sample (SCI, Giovanelli \etal
1997a, 1997b). The line-width dependent scatter of this cluster sample
is well determined.  Since SCI was observed using the same
observational procedures as most galaxies in SFI, the distance
estimates in both samples should suffer from similar observational
uncertainties. However, it is less clear whether the intrinsic scatter
of the TF relation is the same for the field and cluster samples. We
have parameterized the total scatter in the SFI sample by using the
SCI observed scatter and adding an additional intrinsic scatter for
field galaxies in quadrature.  Such a higher scatter for field
galaxies has consistently been found by a number of authors
(e.g. Bothun \& Mould 1987; Freudling, Martel \& Haynes 1991). We
estimated the total scatter of SFI by taking advantage of the distance
dependence of biases in the inferred distances. These biases for field
galaxies are large at high distances and dominate the raw measured
peculiar velocities.  The exact behavior at large distance depends on
the assumed amplitude of the scatter (see Freudling \etal 1995). With
the aid of mock samples, the observed distance dependence of the
average measured peculiar velocity was used to infer the intrinsic
scatter for the SFI sample.

The resulting errors are estimated to be in the range $15-20\%$, and
increasing with decreasing line-width $w$. Following da Costa \etal
(1996), a small fraction ($\sim 7\%$) of galaxies with small
line-width ($\log w \le 2.25$) has been discarded because of the
unreliability of the TF relation and its scatter at such
line-widths. A detailed account on the sample selection, error
estimates, and the procedure of combining the two datasets can be
found in Wegner \etal (1999) and Haynes \etal (1999).  The SCI sample
of $\sim 500$ galaxies within 24 clusters was used for calibrating the
TF relation and in estimating the scatter properties, but the peculiar
velocities of these clusters themselves are not used in this work as
they require a different treatment.

\subsection{Bias Correction}
\label{subsec:bias}

It is crucial to properly correct the data for systematic biases, such
as those arising from the coupling between the random
distance errors, the geometry of space and the inhomogeneities in the
underlying distribution of galaxies, and certain aspects of the sample
selection.  Due to the complexity of the selection criteria and the TF
distance errors in the SFI data, the bias correction could not be
properly estimated using the standard simple analytic expression.
In earlier papers of the SFI series, the bias was
estimated using a numerical Monte-Carlo approach in which
the selection criteria were mimicked in detail (Freudling \etal 1995).
Here, we replace it with a simpler semi-analytic estimate of the bias,
which incorporates the relevant selection criteria.

The bias-correction method will be described in detail by Eldar \etal
(1998). Here we mention only the basic features of the method. Given a
galaxy with a TF inferred distance $d$ and a line-width $\eta=\log
w-2.5$, the Malmquist-corrected distance is adopted to be the
conditional expectation value of the true distance, $r$,
\beq  
E(r|d, \eta) = {\int _0^\infty \,dr\ r\ P(r,d,\eta)\over
                \int _0^\infty dr\, P(r,d, \eta)}, 
\eeq
where $P(r,d,\eta)$ is the joint probability distribution in the
catalog (\eg, Strauss \& Willick 1995).  The line-width is explicitly
included to ensure that the correction holds when the selection
criteria depend on $\eta$.

This joint distribution is derived from several input quantities.  One
is the underlying spatial number density of galaxies, $n(r)$, which is
taken from a self-consistent real-space reconstruction from the IRAS
1.2 Jy redshift survey (as described in Sigad \etal 1998).  Another
input is the distribution of galaxy diameters, $\Phi(D)$.  One also
needs as input the conditional probability $P(\eta,B,I|D)$, that a
galaxy with a given $D$ will have a line-width $\eta$ and absolute
magnitudes $B$ and $I$.  We adopted the same distribution functions as
those used in Freudling \etal (1995). Taking into account the
selection in angular diameter, $a=D/r$, and the apparent blue
magnitude limit $m_{B,{\rm max}}$, one obtains
\beq  
P(r,d,\eta ) \propto
r^2n(r)\int _{-\infty }^\infty da\
S(a|r)\ \Phi(a r)\ r\
\ {\rm exp} \left( -{[{\rm ln}(r/d)]^2 \over 2\Delta^2} \right)\ 
P(r,d,\eta,a r|m_{B,{\rm max}}) . 
\eeq
The selection function of angular diameters at a given true distance,
$S(a|r)$, is derived from the corresponding selection function in
redshift space, $S_z(a|z)$, via
\beq  
S(a|r) = \int _{-\infty }^\infty dz\ S_z(a|z) \ P(z|r), 
\eeq
where $P(z|r)$ is based on the model peculiar-velocity field. 
The joint distribution $P(r,d,\eta,a r|m_{B,{\rm max}})$ is based on a
combination of the diameter--magnitude relation, the correlation
between B and I Magnitudes, the $\eta-B$ relation, and the B-magnitude
limit in the selection of galaxies.

This bias correction scheme was tested, and its details were refined,
using carefully constructed mock catalogs (presented below,
\S~\ref{subsec:mock}). We also tried several variants of the procedure 
to correct the real data for biases.  The results of the
power-spectrum analysis turn out to be fairly insensitive to the
specifics of the bias correction scheme.  In particular, for the
underlying galaxy-density field that enters the correction via $n(r)$,
we tried replacing the IRAS field (Sigad \etal 1998) with a linear
reconstruction of a combination of IRAS and optical data (Freudling,
da Costa \& Pellegrini 1994), and found negligible effects on the
results of the likelihood analysis.

The estimated errors in the observed TF relation can be directly
translated into a distance uncertainty for each galaxy prior to the
correction for biases.  However, the correction for biases changes the
properties of the scatter for a given location in estimated-distance
space, which leads to a different uncertainty in the bias-corrected
distance estimate.  The semi-analytic approach is used also for a
re-evaluation of the distance errors after the bias correction. We
find that the bias correction acts towards slightly decreasing (by
$\sim 10\%$) the average error, because of the additional information
incorporated by the selection effects and the underlying density field
used in the bias correction.  The validity of this approach is
verified using the mock catalogs (which also shows that the
distribution of distance errors after the bias correction closely
resembles a Gaussian distribution, Eldar \etal 1998).  An independent
verification of the magnitude of errors within the framework of the
likelihood analysis is described in \S~\ref{subsec:tlcdmT60}.  In what
follows, we refer to these errors as our `original' error estimates.

\section{METHOD}
\label{sec:method}

\subsection{Likelihood Analysis}
\label{subsec:like}

The goal of this paper is to estimate the power spectrum of mass density
fluctuations from peculiar velocities, by finding maximum likelihood
values for parameters of assumed model power-spectra. Again, the
underlying assumptions are that the velocities and their errors are
Gaussian, and that the velocity correlations can be derived from the
density PS using linear theory. The assumption regarding the
Gaussianity of the velocity field is supported by simulations which
show that it is Gaussian well into the quasi-linear regime (Kofman
\etal 1994). 
This is farther verified for our data set by the fact that the
distribution of observed $\ln(z/d)$ closely resembles a Normal
distribution.  The validity of the second assumption is discussed
later in \S~\ref{subsubsec:nlnl}.  The likelihood analysis method is
described in Zaroubi \etal (1997; see also Kaiser 1988; Jaffe \&
Kaiser 1994).  Here we summarize the main ideas, the underlying
assumptions, and the specific application to peculiar velocities.
Given a data set $\bd$, our objective is to estimate the most likely
model $\bm$. Using Bayes theorem
\beq
\P (\bm \vert \bd ) = {\P(\bm) \P(\bd|\bm) \over \P(\bd)} \,,
\eeq
and assuming a uniform prior $\P(\bm)$,
this can be turned to maximizing the likelihood function, the
probability of obtaining the data given the model, $\L=\P(\bd|\bm)$,
as a function of the assumed model parameters.

Under the assumption that both the underlying velocities and the
observational errors are independent Gaussian random fields, the
likelihood function can be written in the following form
\beq
\label{eq:like}
{\cal L} = [ (2\pi)^N \det(R)]^{-1/2}
  \exp\left( -{1\over 2}\sum_{i,j}^N {u_i R_{ij}^{-1} u_j}\right)\,.
\eeq
This is simply the corresponding multivariate Gaussian distribution,
where $\{u_i\}_{i=1}^{N}$ is the set of $N$ observed peculiar
velocities at locations $\{\br_i\}$, and $R$ is their correlation
matrix. Expressing each data point as the sum of the actual signal and
the observational error $u_i=s_i+\epsilon_i$, the elements in the
correlation matrix have two contributions
\beq
R_{ij}\equiv<u_i u_j>=<s_i s_j>+<\epsilon_i \epsilon_j>=S_{ij}+{\epsilon_i}^2
\delta_{ij}.
\eeq
The first term is the correlation of the signal, that is calculated
from theory. The second term is the contribution of the distance
errors, which are assumed to be uncorrelated. This should be true
for the observational errors and the intrinsic scatter of the TF
relation. We tested the impact of uncertainties in the bias correction, 
which might lead to correlated errors, by varying parameters of our 
bias model within the expected uncertainties. The changes in the results 
reported below are negligible compared to other systematic 
and random errors.  For a given PS, the signal terms are
calculated using their relation to the parallel and perpendicular
velocity correlation functions, $\Psi_{\Vert}$ and $\Psi_{\perp}$,
\beq
S_{ij}=\Psi_{\perp}(r)\sin\theta_i \sin\theta_j + \Psi_{\Vert}(r)\cos\theta_i
\cos\theta_j  \, ,
\eeq
where $r=\vert \br \vert=\vert \br_j-\br_i \vert$ and the angles are
defined by $\theta_i=\hat{\br_i}\cdot\hat{\br}$ (G\'orski 1988; Groth,
Juszkiewicz \& Ostriker 1989).  In linear theory, each of these can be
calculated from the PS,
\beq
\Psi_{\perp,\Vert}(r)= {H_0^2 f^2(\Omega)\over 2 \pi^2}
\int_0^\infty P(k)\, K_{\perp,\Vert}(kr)\, dk \,,
\eeq
where $K_{\perp}(x) = j_1(x)/ x$ and $K_{\Vert}(x) = j_0-2{j_1(x)/
x}$, with $j_l(x)$ the spherical Bessel function of order $l$.  The
cosmological $\Omega$ dependence enters as usual in linear theory via
$f(\Omega)\simeq \Omega^{0.6}$, and $H_0$ is the Hubble constant.

The likelihood analysis is performed by assuming some
parametric functional form for the PS. For each assumed PS, the
correlation matrix $R$ is obtained and used to calculate the
likelihood function (eq.~[\ref{eq:like}]). Exploring the chosen
parameter space, we find the PS parameters for which the likelihood is
maximized.  (Note that since the model parameters appear also in the
normalizing factor of the likelihood function, through $R$, maximizing
the likelihood is {\it not} equivalent to minimizing the $\chi^2$.)
The main computational effort is the calculation and inversion of the
correlation matrix $R$ in each evaluation of the likelihood. It is an
$N \times N$ matrix, where the number of data points $N$ is typically
more than $1000$.

Since the input data are peculiar velocities, the method essentially
measures the combination $f(\Omega)^2P(k)$, and not directly the
mass-density $P(k)$ by itself. This degeneracy between $\Omega$ and the 
PS can be broken when $\Omega$ enters explicitly into the functional form
characterizing the PS shape, as in CDM models (\S~\ref{subsec:models}).

Confidence levels are estimated by approximating $-2\ln\L$ as a
$\chi^2$ distribution with respect to the model parameters.  The
likelihood analysis provides only relative likelihoods of different
models. An absolute measure of goodness-of-fit can be provided, for
example, by the value of the $\chi^2$ obtained with the parameter
values associated with the maximum likelihood.  A $\chi^2$ per degree
of freedom of about unity would indicate that the model provides a
good statistical description of the data.

\subsection{Power Spectrum Models}
\label{subsec:models}

In order to perform the likelihood analysis, a specific parametric
form for the PS is needed. 
For the main analysis of the paper, we use families of 
generalized CDM models normalized by the COBE 4-year data. 
The general form of these models is
\beq
\label{eq:cdm}
P(k) = A_{COBE}(n,\Omega,\Lambda)\,
T^2(\Omega,\Omega_B,h; k)\, k^n\,,
\eeq
where $A$ is the normalization factor and
$T(k)$ is the CDM transfer function proposed by Sugiyama (1995,
a slight modification of Bardeen \etal 1986):
\beq
\label{eq:Tcdm}
T(k) = {\ln\left(1+2.34q) \right) \over 2.34q}
\left[1+3.89q+(16.1q)^2+(5.46q)^3+(6.71q)^4\right]^{-1/4}\,, 
\eeq
\beq
q=k \left[ \Omega h \,
	   \exp (-\Omega_b -(2h)^{1/2} \Omega_b/\Omega)\,  
	   (\ihmpc)  \right]^{-1}\,.
\eeq
 
These models include 
open universes
with no cosmological constant,
flat models with a cosmological constant ($\Omega+\Omega_\Lambda=1$),
and tilted models with a large-scale power index $n$ that can be
different from unity. The latter may include tensor fluctuations with
tensor to scalar ratio of quadrupole moments of $T/S=7(1-n)$.  The
free parameters in the CDM models are $\Omega$, h and n.  In all cases
the baryonic density is set to be $\Omega_b=0.024 h^{-2}$ (\eg,
Tytler, Fan \& Burles 1996).  For each model, the amplitude A is fixed
by the COBE 4-year data.

We followed the COBE normalization adopted in Zaroubi \etal (1997), who
used the COBE DMR data (Hinshaw \etal 1996) to set the PS
amplitude calculated by different authors (G\'orski \etal 1995; Sugiyama
1995; White \& Bunn 1995) for various cosmological CDM-like models.
The calculation of Sugiyama (1995) was used as a reference. For
models not studied by him the other results were used, after matching
them to Sugiyama's in the commonly studied models. For a summary of
the COBE normalization results see G\'orski \etal (1998).
 
In addition, we use a different parameterization of the same power spectra, 
namely 
\beq
\label{eq:gamma}
P(k)= A\, k\, T^2(k), \quad
T(k) = \Bigl( 1 + [ ak/\Gamma + (bk/\Gamma)^{3/2} + (ck/\Gamma)^2 ] ^{\nu}
\Bigr)^{-1/\nu}\,,
\eeq
with $a=6.4\hmpc$, $b=3.0\hmpc$, $c=1.7\hmpc$ and $\nu = 1.13$ (\eg,
Efstathiou, Bond \& White 1992).  In the context of the CDM model,
$\Gamma$ has a specific cosmological interpretation, $\Gamma=\Omega
h$. Below, however, we use equation (\ref{eq:gamma}) as a generic
form with limiting logarithmic slopes $n=1$ and $-3$ on large and
small scales respectively, and with a turnover at some intermediate
wavenumber that is determined by the single shape parameter $\Gamma$.
Hereafter, we refer to this functional form of the power-spectrum as
the ``$\Gamma$ model''. We use it as a convenient parameterization,
for comparability with other works, and for relaxing the COBE
normalization.  The free parameters that we vary in this case are the
amplitude $A$ and the shape parameter $\Gamma$.

\subsection{Error Models}
\label{subsec:errors}

We make a special effort to estimate the distance errors.  As
mentioned in section~\ref{sec:intro}, this is done because the
amplitude of the deduced PS depends on their sum in quadrature; if
errors are overestimated, the PS is underestimated and vice versa.

We first apply the likelihood analysis with the original distance
errors, $\sigma_{oi}$, as estimated a priori for each galaxy ($i$) in
the SFI catalog with the procedure explained in
section~\ref{subsec:bias}.  Alternatively, we incorporate the errors
in the likelihood analysis itself, by allowing a parametric model for
the errors in addition to the parametric model of the PS. An error
model is fully specified by the standard deviations $\sigma_i$ because
we assume that the distance errors for the individual galaxies are
uncorrelated and that the scatter is Gaussian. We try two alternative
global modifications of the original errors as our error model: one is
based on a free multiplicative factor, $\sigma_i = p \sigma_{oi}$, and
the other is based on a free additive constant in quadrature,
$\sigma_i = (\sigma_{oi}^2 \pm q^2)^{1/2}$. The latter is similar to
the way we modeled the difference in scatter between the field and
cluster samples (see \S~\ref{subsec:derrors}).  The errors are
incorporated in the model that constitutes the correlation matrix, and
the parameters $p$ or $q$ are adjusted simultaneously with the
parameters of the PS until the likelihood is maximized.

The apparent cost of adding the error parameter to the likelihood
analysis is a larger formal error in the final results for the power
spectrum and the cosmological parameters.  However, since our original
error estimates carries some uncertainty, this procedure, which
provides an almost independent estimate of the errors, could add to
the overall confidence in our results.
   
\section{TESTING THE METHOD}
\label{sec:mock}

\subsection{Mock Catalogs}
\label{subsec:mock}

It is essential to check the method with realistic mock catalogs, in
view of the large errors in the data and the approximations made in
the analysis.  For this purpose, we use the $N$-body simulation of
Kolatt \etal (1996) which was designed to mimic the large-scale
density distribution in our local universe. The simulation is based on
initial conditions extracted from a reconstruction of the smoothed
($5\hmpc$ Gaussian) real-space density field from the IRAS 1.2 Jy
redshift survey, taken back into the linear regime.  Small-scale
perturbations were added by means of constrained random realizations,
and the system was then evolved forward in time using a particle-mesh
$N$-body code until a present epoch defined by $\sigma_8=0.7$.  The
``true'' PS was calculated directly from the underlying mass
distribution of the simulation, by Fourier transforming to k-space and
calculating the power in bins of wavelength.

``Galaxies" were identified in the simulation via a linear biasing
scheme, and then divided into galaxy types, S's and E's, while obeying
the morphology-density relation.  Observational parameters were
assigned to the S galaxies in the mock sample according to the
prescription of Freudling \etal (1995), and perturbed at random
according to the estimated observational errors. Subsequently, we
selected ten random mock SFI samples using the exact selection
criteria of the real SFI sample. Each of these mock catalogs was
corrected for biases, and the errors were re-evaluated accordingly, in
the same way as in the real data (see Eldar \etal 1998).

\subsection{Testing with the $\Gamma$ Model}
\label{subsec:mock_gamma}

We first apply the likelihood analysis to the mock SFI catalogs using
the $\Gamma$ functional form (eq.~[\ref{eq:gamma}]) as the prior model
for the PS.  We allow the amplitude $A$ and the shape parameter
$\Gamma$ to vary, and include an additional free parameter in the
error model. It is realized that the freedom provided by this family
of models (just as by any other family of models) may not be enough
for an adequate fit to the true PS.  No additional constraint is
applied on large scales, so this is a test of the ability of the
velocity data alone to constrain the PS.

\begin{figure}[tbp]
{\includegraphics{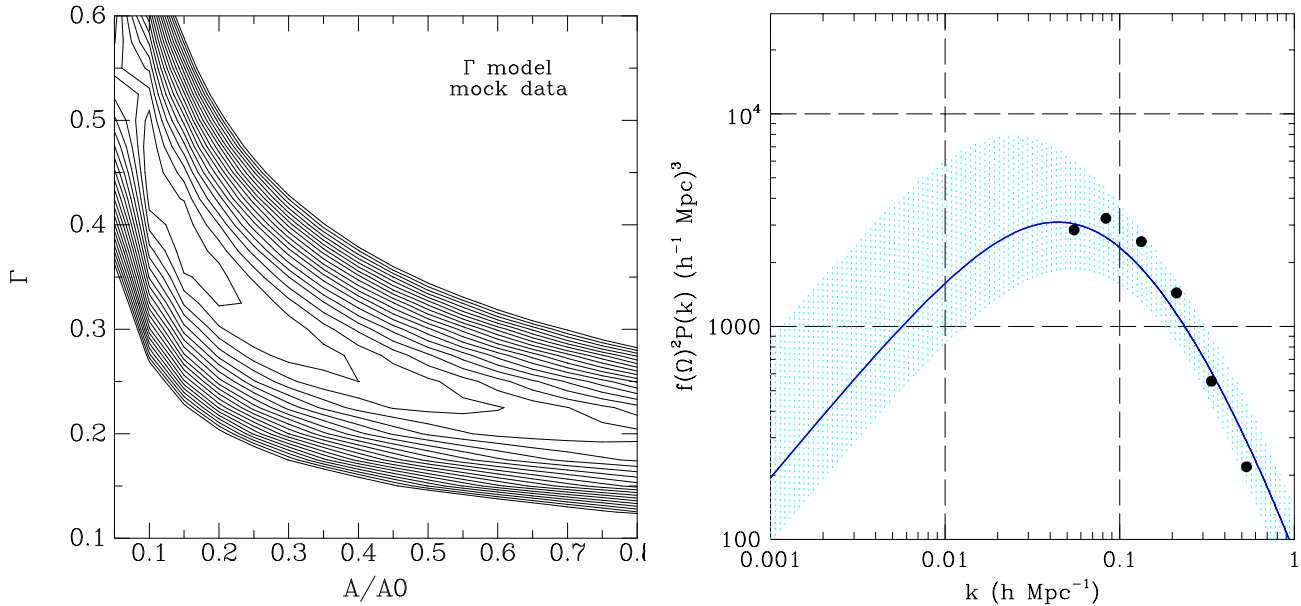}}
\vspace{7.7 cm}
\caption{\capt
Left panel: Contour map of $\ln \L$ in the $A-\Gamma$ plane for one
random mock catalog. Contour spacing is $\Delta[\ln \L]=-1$. $A$ is in
units of $A_0 = 2.0 \times 10^6 (\hmpc)^4$.  The errors were varied by
a multiplicative factor. Right panel: The PS corresponding to the
maximum-likelihood $\Gamma$-model parameters determined for the mock
catalog. The filled symbols mark the true PS of the simulation. The
shaded area around the PS is the $90\%$ confidence region for the
best-fit errors, obtained from the contour map. }
\label{fig:mockg}
\end{figure}

Figure~\ref{fig:mockg} (left panel) shows a contour plot of the
resulting log-likelihood ($\ln \L$) in the parameter plane
($A-\Gamma$), as obtained from one of the realizations of the mock
catalogs.  The errors in this case were allowed to vary by the
multiplicative factor $p$, and the plot shown corresponds to the
best-fit error parameter.  Here, and in all the figures that follow,
the log-likelihood contours are relative to the maximum likelihood
with contour spacing of $\Delta[\ln \L]=-1$.  The right panel of
Figure~\ref{fig:mockg} shows the corresponding best-fit power spectrum
(solid line). The filled symbols mark the target of the reconstruction
--- the true PS of the simulation.  The shaded area about the derived
PS corresponds to the region of $90\%$ confidence about the
most-likely parameters in the likelihood plot, for fixed errors. The
uncertainty becomes large at small $k$'s corresponding to scales
larger than the sampled volume, because no additional data were used
to constrain the PS on large scales.  The figure demonstrates that for
this random realization the likelihood analysis with the $\Gamma$
model recovers the true PS well within the error-bars.  A similar
quality of recovery is obtained for all the random realizations of the
mock catalogs, and also when the errors are varied in the alternative
way.

The maximum-likelihood errors are found to be within $5\%$ of their
``true" values. The latter were estimated by slightly modifying the
known distance errors (as built into the mock catalogs) after
correcting for Malmquist bias. The $5\%$ error reflects the imperfect
match between the assumed family of shapes for the PS and the true
shape, and, perhaps, the uncertainty in the modification of the error
estimate or the slight deviation of the modified errors from a
Gaussian distribution.

\subsection{Testing with a Tilted \lcdm  Model}
\label{subsec:mock_tlcdm}

We wish to check the success of the likelihood analysis also with the
COBE-normalized CDM models. We choose as our test case the flat
($\Omega+\Omega_\Lambda=1$) \lcdm family of models, with tensor
fluctuations, a corresponding tilt in $n$, and a Hubble constant of
$h=0.6$.  COBE normalization is imposed as if the mock simulation is
identical to the real universe. The likelihood analysis is thus
performed by varying the parameters $\Omega$, $n$ and the
error-parameter $p$ or $q$.  This family of shapes for the PS is,
again, not guaranteed to provide a perfect fit to the true PS. In
particular, the parameter-dependent COBE normalization is not
guaranteed to give the correct amplitude, since the simulation was not
explicitly constrained to produce the level of large-scale CMB
anisotropies detected in the real universe.

\begin{figure}[tbp]
{\includegraphics{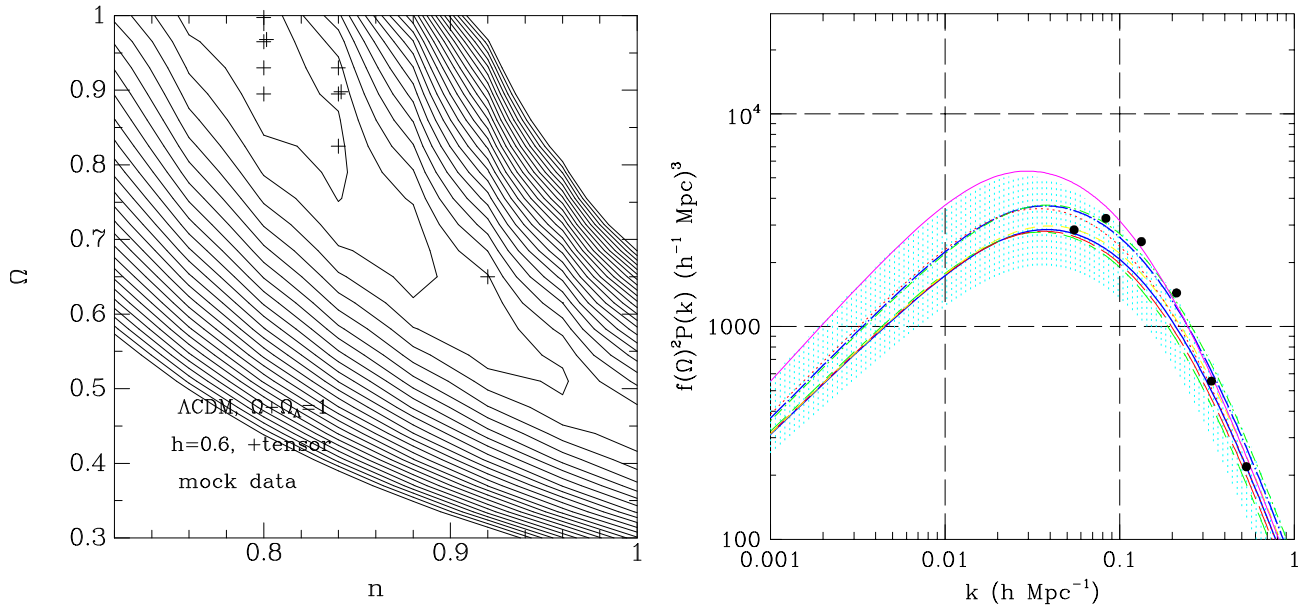}}
\vspace{7.7 cm}
\caption{\capt
Left panel: Contour plot of $\ln\L$ in the $n-\Omega$ plane for one of
the mock catalogs, for the tilted \lcdm model with tensor fluctuations
and $h=0.6$, and the errors varied in quadrature.  The best-fit values
for all catalogs are marked by `+'.  Right panel: Best-fit PS of the
10 mock catalogs (thick lines representing same curves derived for two
different catalogs). The shaded area represents the $90\%$ confidence
region for the catalog whose contour plot is shown.  The filled
symbols mark the true PS of the simulation. }
\label{fig:mockt}
\end{figure}

Figure~\ref{fig:mockt} (right panel) shows the best-fit power spectra
of the 10 mock SFI catalogs, superimposed on the true PS. This test
uses the $q$ error parameter. The left panel shows $\ln \L$ contours
in the $\Omega-n$ plane for one representative mock catalog, with the
maximum-likelihood points for all ten catalogs marked. We see that all
the maximum-likelihood points fall along the ridge of high-likelihood
in the one case plotted, and are therefore moderately consistent with one
another. A way to translate the likelihood contours to errors in the
values of the model parameters is by assuming that, with the errors
fixed, $-2\ln\L$ has a $\chi^2$ distribution with two degrees of
freedom. Then, the $1\sigma$ confidence level around the
maximum-likelihood point is at $\ln\L \sim -1.15$ and the $90\%$
confidence level is at $\ln\L \sim -2.3$. The fact that indeed six of
the ten cases fall within the $1\sigma$ contour as determined above,
and nine cases fall within the $90\%$ confidence level, indicate that
this crude error estimate is quite reasonable. The $90\%$ confidence
region for this specific catalog is again drawn as a shaded area in
the PS plot; one can see that this region indeed resembles the actual
scatter of the ten cases.

The maximum-likelihood power-spectra fit reasonably well the true PS,
with a fairly small spread on small and intermediate scales. For large
scales (small $k$'s) the scatter is somewhat larger, but not as large
as for the $\Gamma$ model which was completely free at large
scales. Again, the success of recovery is similar when the alternative
error parameter is used, and the errors are similar to those obtained
in the case of the $\Gamma$ model.  It is encouraging to note that the
recovery of the PS is fairly robust on the relevant scales ($k\sim
0.1\ihmpc$) among the realizations, and independent of the prior model
assumed for the PS, or the assumed error model.

\section{RESULTS}
\label{sec:results}

\subsection{Maximum-Likelihood Errors}
\label{subsec:tlcdmT60}

Before estimating the PS from the actual SFI data, we investigate the
reliability of our observational error estimate by allowing certain
freedom in the errors. As a test case we use as a prior for the PS the
COBE-normalized \lcdm family of models, with tensor fluctuations and a
corresponding tilt in $n$, and with the Hubble constant fixed at
$h=0.6$.  We perform the likelihood analysis on the real SFI data
varying $\Omega$ and $n$, with the errors treated in three different
ways; first with the errors fixed at their original values, and then
by varying them according to the two error models discussed above.
   
\begin{figure}[t]
{\includegraphics{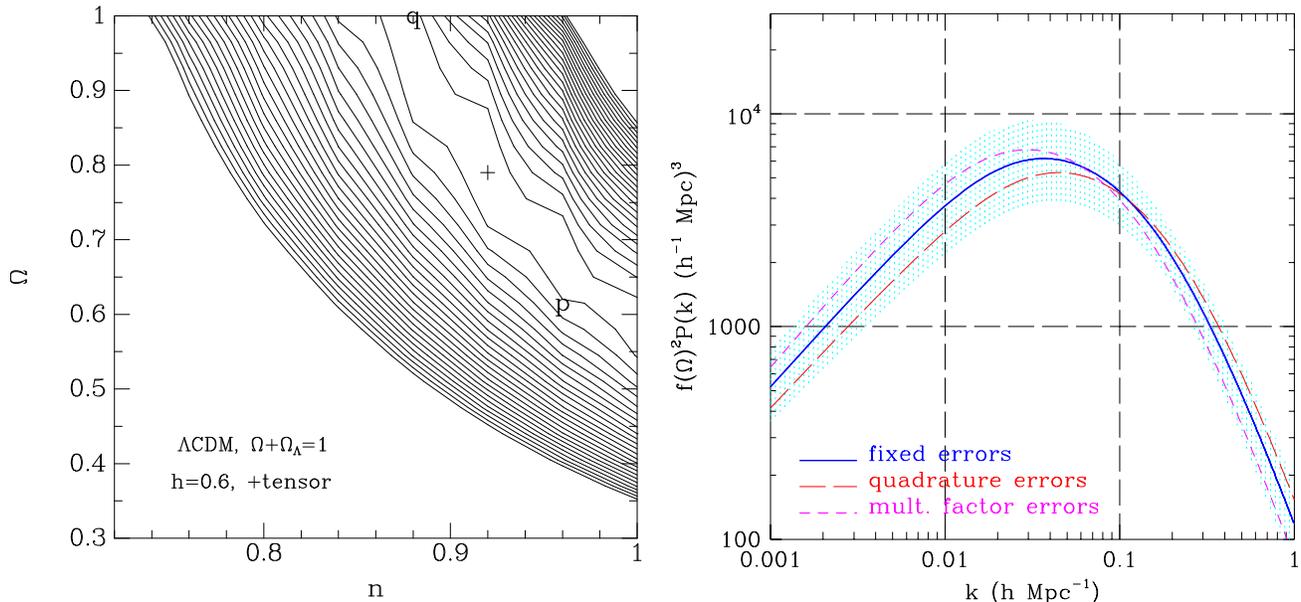}}
\vspace{7.6 cm}
\caption{\capt
Left panel: Contour plot of $\ln\L$ in the $\Omega-n$ plane for the
SFI sample, for the tilted \lcdm model with tensor fluctuations and
$h=0.6$, using the original error estimates. The best-fit point is
marked with a `+'. The maximum likelihood locations when varying the
errors in quadrature (`$q$') or by a multiplicative factor (`$p$') are
also marked. Right panel: The most-likely PS for this model, for these
three variants of the errors. The dotted region around the PS
represents the $90\%$ confidence limit for the case of the original
errors (thick line), obtained from the high-likelihood ridge shown in
the contour map.  }
\label{fig:tlcdmT60}
\end{figure}

Figure~\ref{fig:tlcdmT60} summarizes the results obtained in these
cases.  The left panel shows the $\ln-$likelihood contours in the
$\Omega-n$ plane for the case of fixed errors, with the best-fit
points for the three cases marked. The corresponding power spectra are
presented in the right panel.  In the $p$ case, the preferred errors
are $5\%$ larger than the original ones, while in the $q$ case the
preferred errors are smaller by 0.03 in quadrature (typically a
decrease of $\sim 2\%$).

The different trends in the likely errors reflect our uncertainty of
the exact form of the error model.  We note that while these changes
are in opposite directions, they are of small magnitude, within the
uncertainty expected based on the mock catalogs. The corresponding
changes in the best-fit parameters are along the ridge of high
likelihood in the $\Omega-n$ plane, within the $1\sigma$ confidence
level, \ie, it is hardly significant.  In all three cases,
$\chi^2/N_{dof} \sim 1$ for the best-fit PS ($1.02$ for the original,
fixed errors, $0.99$ for the p error model and 1.02 for the q error
model), implying that all are reasonable fits to the data.  Similar
results concerning the errors are obtained when the other PS models
are used as priors. The error estimate is robust to variations in the
original errors about which the error model is perturbed. This
likelihood analysis of the errors thus provides a very encouraging
indication that the original error estimates in SFI are accurate to
better than $5\%$.  Note that ``original'' here refers to the refined
SFI errors after the correction for biases (\S~\ref{subsec:bias}). The
fact that the likelihood analysis and the semi-analytic correction
converge to the same error estimate is  encouraging. Based on
this finding, we perform the rest of the analysis in this paper using
fixed errors at their original values.

\subsection{COBE Normalized CDM Models}
\label{subsec:cdm}

We now use the generalized CDM families of cosmological models of the
form described in equations (\ref{eq:cdm}) and (\ref{eq:Tcdm}).  Our
models include open CDM (\ocdm), flat models with a cosmological
constant, and tilted models with or without tensor fluctuations,
allowing for variations in the cosmological parameters $\Omega$, $h$,
and $n$.  For each specific choice of model and parameters the
amplitude is fixed according to the 4-year COBE normalization.

\subsubsection{Scale-Invariant Models}
\label{subsubsec:si_cdm}

\begin{figure}[tbp]
{\includegraphics{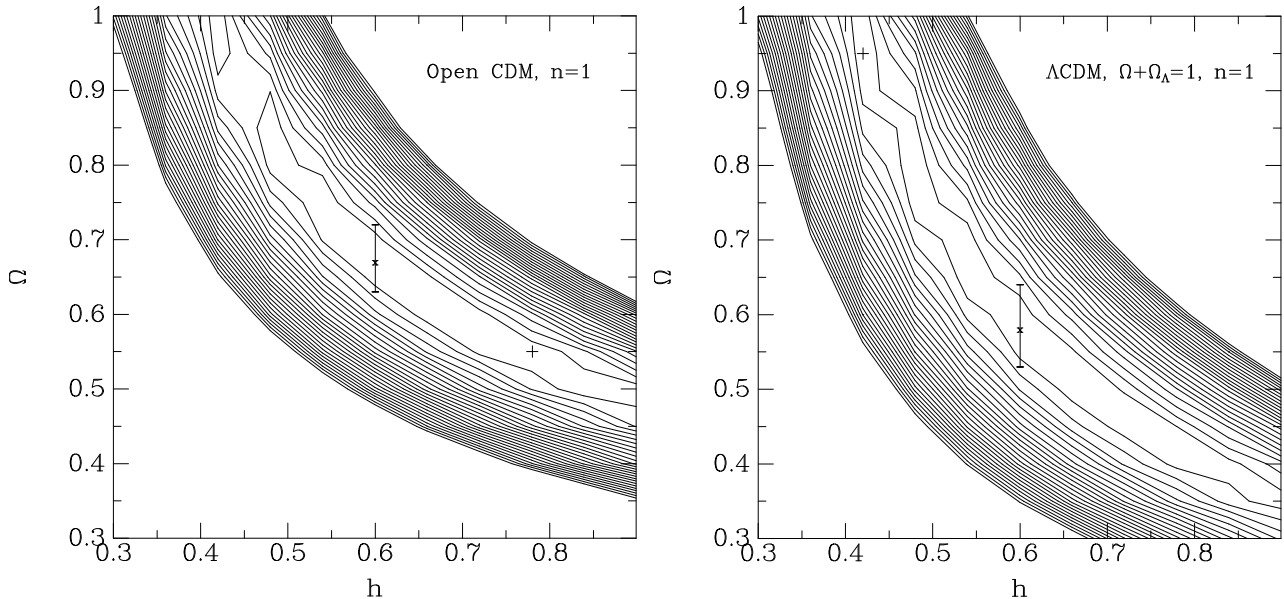}}
\vspace{7.6 cm}
\caption{\capt
Contour plot of $\ln\L$ in the $\Omega-h$ plane for scale-invariant
\ocdm model (left panel) and \lcdm model (right panel). The
most-likely value of $\Omega$ and its $90\%$ error bar are marked for
a fixed value of $h=0.6$.  }
\label{fig:olcdm}
\end{figure}        

Figure~\ref{fig:olcdm} shows the resulting likelihood contours for the
scale-invariant case, $n=1$, for the \ocdm model and the \lcdm model.
The contours are plotted in the $\Omega-h$ plane.  The best-fit
parameters in each case are marked, but it is clear from the elongated
contours that the two parameters are not determined separately. The
high-likelihood ridges rather constrain a degenerate combination of
these parameters, which can be roughly fitted by the following
functions:
\begin{eqnarray}
\label{eq:ocdm}
\Omega\, {\h60}^{0.9} & = & 0.68\pm 0.06\,, \qquad \rocdm; \\
\label{eq:lcdm}
\Omega\, {\h60}^{1.3} & = & 0.59\pm 0.07\,, \qquad  \rlcdm \,.
\end{eqnarray}
The error-bars (here, and throughout the paper) arise from the joint
$90\%$ confidence region of the parameters.  
The constraints on $\sigma_8 f(\Omega)$, obtained by integrating over
the corresponding power spectra, are ${0.83}^{+0.07}_{-0.11}$ and
${0.81}^{+0.13}_{-0.07}$ for \ocdm and
\lcdm respectively. 
The error-bars quoted are the marginalized 1-dimensional $90\%$ confidence 
limits.
For an assumed value of $h$, \eg, $h=0.6$, the maximum-likelihood values 
are $\Omega=0.67 \pm 0.05$ for \ocdm and $\Omega=0.58 \pm 0.06$ for \lcdm 
(as marked on the plots).  
The $\chi^2/N_{\rm dof}$ for the best-fit PS are 
$1.01$ and $1.04$ respectively, with similar values along the
high-likelihood ridge.  With $N_{\rm dof}=1213$, one expects for a
good fit $1.00\pm0.04$, so our CDM models indeed provide good fits to
the data.

\subsubsection{Tilted Models}
\label{subsubsec:tcdm}

\begin{figure}[tbp]
{\includegraphics{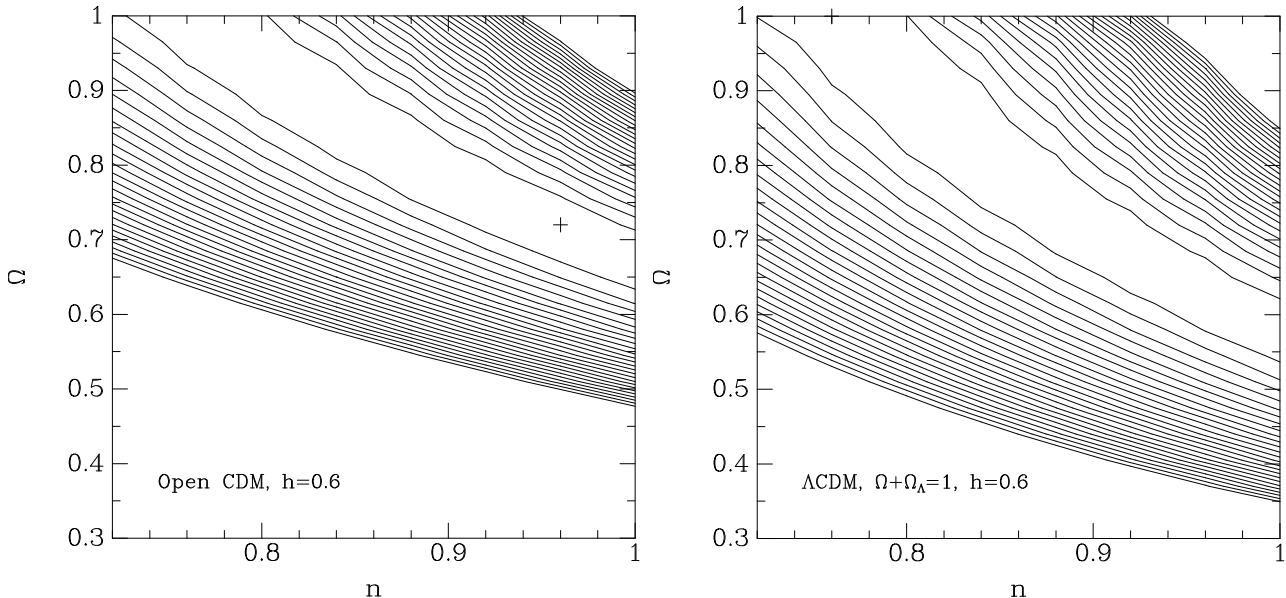}}
\vspace{7.6 cm}
\caption{\capt
Contour plot of $\ln\L$ in the $\Omega-n$ plane, for the tilted \ocdm
model (left panel) and tilted \lcdm model (right panel). In both cases
$h=0.6$ and no tensor component is included.  }
\label{fig:tolcdm}
\end{figure}

Figure~\ref{fig:tolcdm} presents the results obtained when allowing
for a tilt in the PS on large scale relative to $n=1$, both for the
\ocdm and \lcdm families of models. The first cases considered are
with scalar fluctuations only, $T/S=0$.  We fix the Hubble constant
here at $h=0.6$ while varying $\Omega$ and $n$.  Again, the elongated
ridge of high-likelihood determines a certain degenerate combination
of the parameters, which can be approximated by:
\begin{eqnarray}
\label{eq:tocdm}
\quad \Omega \, n^{1.4}\,({\h60}^{0.9})& = &0.68\pm 0.07\, ,\qquad\rocdm ;\\
\label{eq:tlcdm}
\quad \Omega \, n^{2.0}\,({\h60}^{1.3} )& = &0.58\pm 0.08\, ,\qquad\rlcdm \, . 
\end{eqnarray}
The $h$ dependence is determined for the $n=1$ case. The corresponding
constraints are $\sigma_8 f(\Omega)={0.83}^{+0.08}_{-0.10}$ for the
tilted \ocdm case and $\sigma_8 f(\Omega)={0.82}^{+0.10}_{-0.09}$ for
tilted \lcdm.  The $\chi^2/N_{\rm dof}$ values are $1.02$ in both
cases, again a good fit.

The case of the tilted \lcdm family of models, with $h=0.6$ and with a
tensor component of $T/S=7(1-n)$, was partly discussed already as our
default case in \S~\ref{subsec:tlcdmT60}. The likelihood map in
Figure~\ref{fig:tlcdmT60} reveals the familiar situation of a
high-likelihood ridge that constrains a degenerate combination of the
cosmological parameters, now approximated by
\begin{eqnarray}
\label{eq:tlcdmT60}
\qquad \Omega\, n^{3.9} ({\h60}^{1.3})& = &0.58\pm 0.08\,, \qquad  
\rlcdm\,+{\rm tensor}\,.
\end{eqnarray}
The $h$ dependence is determined for $n=1$.  The corresponding value
of $\sigma_8 f(\Omega)$ is $0.81^{+0.09}_{-0.08}$.  The uncertainty
associated with the PS, shown as the shaded area in the right panel of
Fig.~\ref{fig:tlcdmT60}, is similar to the uncertainty in the other
COBE-normalized CDM variants.

\subsection{$\Gamma$ Model}
\label{subsec:gamma}

\begin{figure}[t]
{\includegraphics{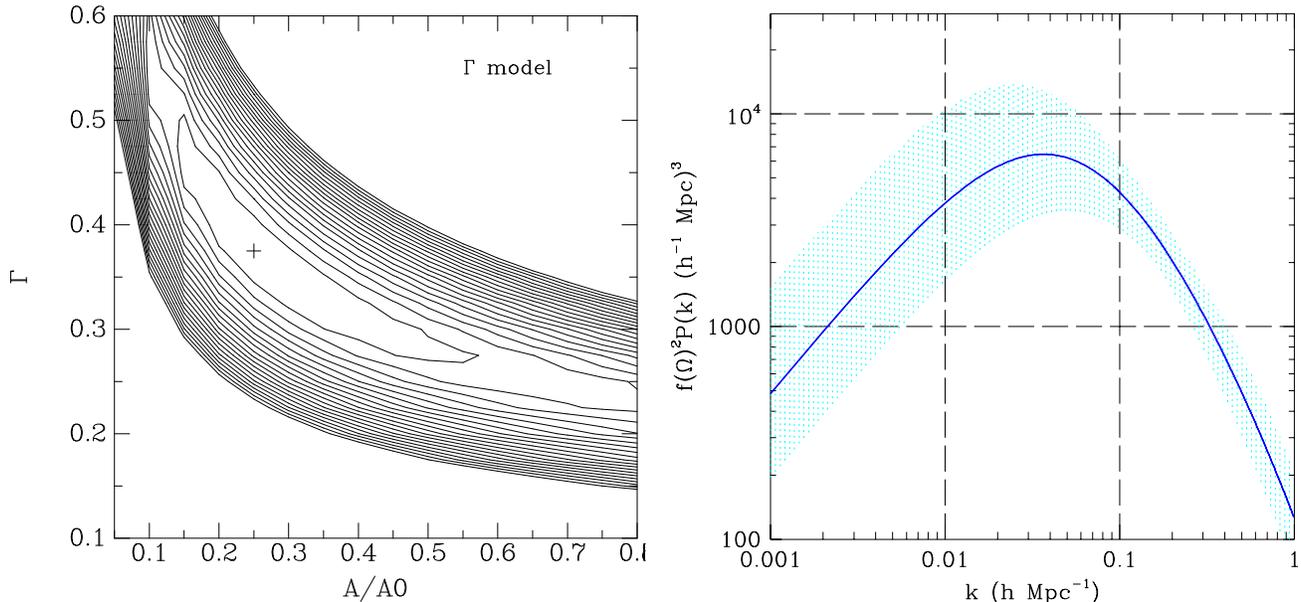}}
\vspace{7.6 cm}
\caption{\capt
Left panel: Contour plot of $\ln\L$ for the $\Gamma$ model. The
best-fit point is marked with a `+'. Right panel: The best-fit PS,
with the shaded area marking the uncertainty based on the $90\%$
confidence region of the likelihood contours.  }
\label{fig:gamma}
\end{figure}

Finally, we use the $\Gamma$ model as a prior for the PS, varying the
amplitude $A$ and shape parameter $\Gamma$ with no additional
constraints imposed at large scales.  Figure~\ref{fig:gamma} shows the
contours of $\ln \L$ in the $A-\Gamma$ plane, and the corresponding
best-fit PS. The maximum likelihood values are $\Gamma=0.375 \pm 0.14$
and $A = 5.0\times 10^5 (\hmpc)^4$.  The $\chi^2$ per degree of
freedom for the maximum likelihood parameters is $\chi^2/N_{\rm
dof}=1.03$, indicating that this is a good fit to the data.  The
constraint obtained by integrating over the power spectra is $\sigma_8
f(\Omega) = {0.80}^{+0.09}_{-0.08}$.  The scatter at small $k$'s is
larger than in the case of the COBE-normalized models, due to the
amplitude freedom, as seen already in the mock catalogs.

\section{ROBUSTNESS OF RESULTS}
\label{sec:robust}

The error estimates in the parameters given in the previous section
are formal $90\%$ confidence levels. In this section, we test the
robustness of these results to various variations in the data and
models used.

\subsection{Robustness to Models}
\label{subsec:robust}

\begin{figure}[tbp]
{\includegraphics{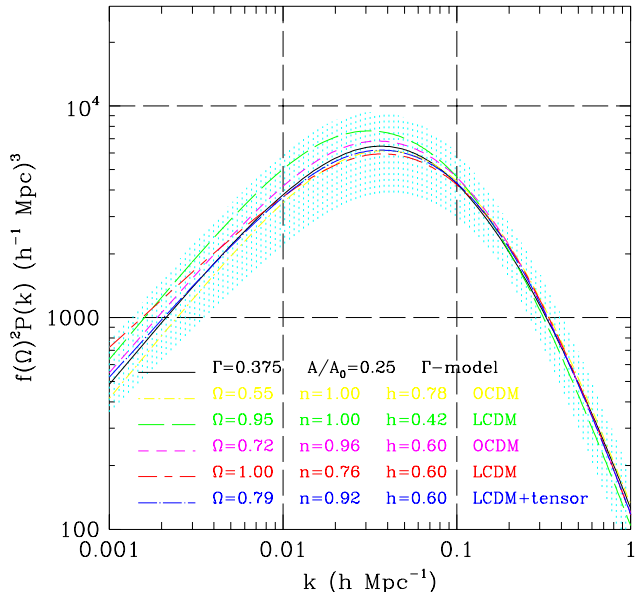}}
\vspace{7.6 cm}
\caption{\capt
The maximum-likelihood power spectra based on the $\Gamma$ model
(solid line) and the various COBE-normalized CDM models. The shaded
area is once again the $90\%$ confidence region for the tilted \lcdm
model with tensor fluctuations and $h=0.6$.  }
\label{fig:all}
\end{figure}

Figure~\ref{fig:all} shows the power spectra corresponding to the
maximum-likelihood parameters for all the models presented so far in
this paper, including the COBE-normalized CDM variants and the
$\Gamma$ model.  The $90\%$ confidence region for the tilted \lcdm
model with tensor fluctuation and $h=0.6$ is drawn as well, as a
reference for the uncertainty associated with each model based on the
likelihood contours.  The similarity of all the curves is striking;
they agree well within the formal uncertainties of each other. The
agreement is excellent for $k>0.1$, on the scale where the data
constrain the models effectively.  The difference between the curves
shows a slightly larger scatter on larger scales, not properly sampled
by the present data. The similarity of the results using as priors the
COBE-normalized CDM models and the amplitude-free $\Gamma$ model
indicates that the peculiar velocity data themselves contain
meaningful information to constrain the PS.

Table~1 summarizes the features of the most likely power spectra based
on the various prior models. The approximate constraint on the
combination of cosmological parameters as obtained from the
high-likelihood ridge is given for each case. The best-fit values of
the individual cosmological parameters are also listed, but recall
that they carry large uncertainties.  The exact location of the
maximum-likelihood point in the high-likelihood ridge is hardly
significant. Parameters that were held fixed in the likelihood
analysis are marked in brackets. Several characteristics of the
best-fit power spectra are listed: the value of $\sigma_8 f(\Omega)$,
the amplitude of $f(\Omega)^2 P(k)$ at $k=0.1 \hmpc$ and the location
of the PS peak, $k_{peak}$. The error-bars quoted in the header
represent the typical $90\%$ confidence uncertainty in these
quantities within each family of models.

\begin{table}[tb]
\label{table}
\caption{Maximum-Likelihood Results for the various models}
\vspace{0.5 cm}
\begin{tabular}{@{}c@{ }c@{}c@{}c@{}c@{}c@{ }@{ }c@{ }@{ }c@{}c@{ }c@{}} 
\multicolumn{10}{c}{COBE-normalized CDM models}\\
\hline \hline
\vspace{0.2 cm}
  &  & $\sigma_8\Omega^{0.6}$ & $P_{0.1}\Omega^{1.2}$ & $k_{peak}$ &  &
&  &  & \\ 
  & High Likelihood &  & \small{($h^{-3}Mpc^3$)} & \small{($hMpc^{-1}$)} 
& & & & & \\ 
 CDM Model & Ridge & \small{($\pm0.10$)} & \small{($\pm1500$)} & 
\small{($\pm0.01$)} &
$\Omega$ & $n$ & $h$ & $\chi^2/N_{\rm dof}$ & $-\ln{\cal L}$ \\ \hline

\vspace{0.4 cm}
 Open, n=1  & $\Omega {\h60}^{0.9}=0.68\pm 0.06$ & 0.83 & 4400 & 
0.038 & 0.55 & (1) & 0.78 & 1.01 & 8579.4 \\ 

\vspace{0.4 cm}
 $\Lambda$, n=1 & $\Omega{\h60}^{1.3}=0.59\pm 0.07$ & 0.81 & 4600 & 
0.031 & 0.95 & (1) & 0.42 & 1.04 & 8580.1 \\  

\vspace{0.4 cm}
 Tilted-Open & $\Omega n^{1.4}=0.68\pm 0.07$ & 0.83 & 4600 & 
0.035 & 0.72 & 0.96 & (0.6) & 1.02 & 8579.5 \\
 
\vspace{0.4 cm}
 Tilted-$\Lambda$ & $\Omega n^{2.0}=0.58\pm 0.08$ & 0.82 & 4200 &
0.037 & 1.00 & 0.76 & (0.6) & 1.02 & 8579.6 \\            

 Tilted-$\Lambda$ & $\Omega n^{3.9}=0.58\pm 0.08$ & 0.81 & 4300 &
0.037 & 0.79 & 0.92 & (0.6) & 1.02 & 8579.5 \\

 +tensor & & & & & & & & & \\

\hline 
 $\Gamma$ model & $\Gamma=0.375\pm 0.14$ & 0.80 & 4300 & 0.037 & & & & 
1.03 & 8579.3 \\

\end{tabular}
\end{table}

The typical results for the PS are $P(k=0.1 \ihmpc) \Omega^{1.2}
=(4.4\pm1.5)\times 10^3 \3hmpc$ and $\sigma_8 \Omega^{0.6} = 0.82 \pm
0.10$.  The variations from model to model are much smaller than the
formal errors for each model, increasing the above errors to $1.7$ and
$0.12$ respectively.  These results are thus almost independent of the
model, at least for the family of models considered here. The actual
likelihood values of all the best-fit models are very similar,
and all have comparable $\chi^2/N_{dof} \simeq 1$ values.  The
variation of the high-likelihood ridge between the \lcdm and \ocdm
families of models is more noticeable.  The general constraint on the
combination of cosmological parameters can be roughly approximated by
$\Omega\, n^{\nu}\, {\h60}^{\mu} = 0.62 \pm 0.15$, where the error
includes the formal uncertainties of the three parameters and the
variations between models. For \lcdm, $\mu=1.3$ and $\nu=2.0,3.9$
without and with tensor fluctuation respectively. For \ocdm, without
tensor fluctuations, the powers are $\mu=0.9$ and $\nu=1.4$.

The similarity of the power spectra obtained using 
the COBE-normalized CDM models and the COBE-free $\Gamma$ model (see
Table~1) indicates that the PS is predominantly determined by the
velocity data.  Therefore, we have so far ignored the error associated
with the COBE normalization.  As a test for the sensitivity to this
error, we have repeated the analysis using the 
tilted \lcdm model (with tensor fluctuations),
but now normalized alternatively $\sim 18\%$ higher or lower then the 
mean COBE values (in accordance with the relative $\pm 1 \sigma$ uncertainty 
associated with $Q_{rms-PS \vert n=1}$, Bennett \etal 1996).
This results in a slight shift of the high-likelihood ridge,
corresponding to a $\sim 6\%$ change in the constraint on the 
combination of parameters (eq.~[\ref{eq:tlcdmT60}]; a smaller value is 
obtained for the higher normalization and vice versa),
which is within our formal $1 \sigma$ error-bars. 
However, the combined effect of the different amplitude and
corresponding cosmological parameters on the PS is essentially
negligible, with $\sigma_8 \Omega^{0.6}$ varying by only $0.01$.

\subsection{Zero-point Uncertainty}
\label{subsec:0p}

\begin{figure}[t]
{\includegraphics{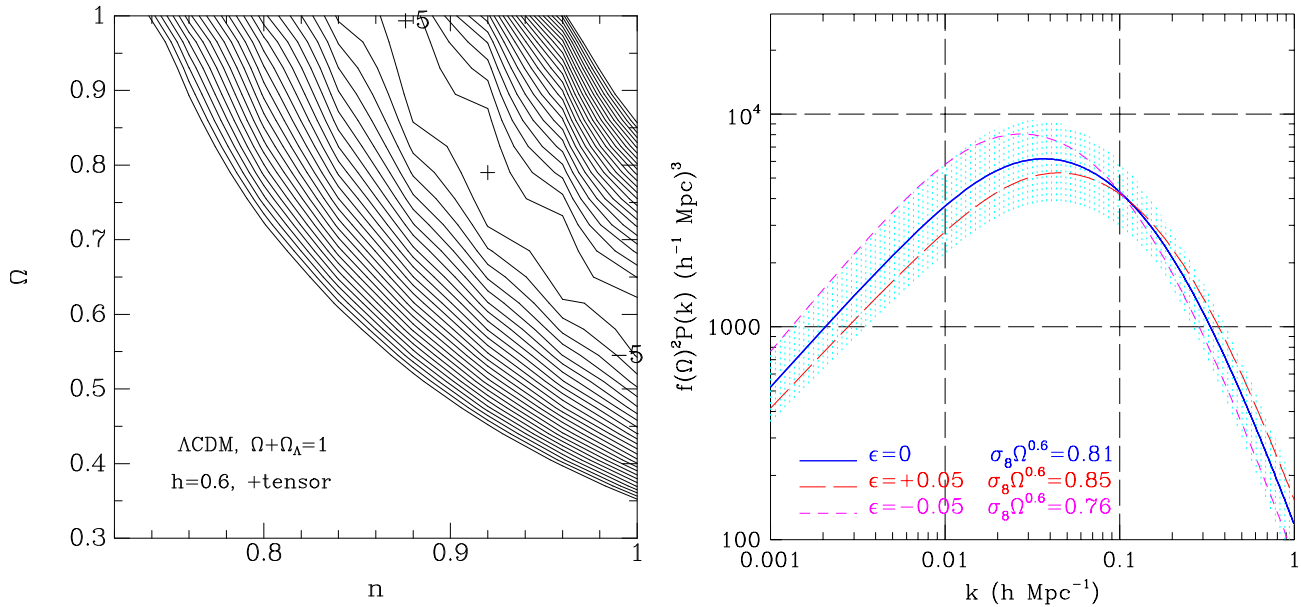}}
\vspace{7.7 cm}
\caption{\capt
Left panel: Contour plots in the $\Omega-n$ plane for the tilted \lcdm
$h=0.6$ model with a tensor component, for the original
calibration. The maximum-likelihood point is marked by `+'. The
maximum-likelihood values when varying the global zero-point by a $\pm
5\%$ Hubble flow are marked by `+5',`-5' accordingly.  Right panel:
The most likely PS for the original zero-point calibration (solid
line) and when varying the zero-point by $\epsilon=+0.05$ (long dashed
line) and $-0.05$ (short dashed line). The corresponding $\sigma_8
\Omega^{0.6}$ values are marked on the plot. The shaded region is the
formal likelihood $90\%$ confidence region for the original case.  }
\label{fig:0point}
\end{figure}        

A fundamental freedom in the measured peculiar velocities is in the
global zero-point of the TF relation, which fixes the distances at
absolute values (in $\kms$).  Changing the zero-point, that is
multiplying the distances $r$ by a factor $(1-\epsilon)$, is
equivalent to adding a monopole Hubble-like flow $\epsilon r$ to the
peculiar velocities. The zero-point calibration of the TF relation
used for the SFI sample was obtained from the SCI catalog of $\sim
500$ galaxies within $24$ clusters, using the ``Basket of Clusters''
approach (Giovanelli \etal 1997a, 1997b). The uncertainty in the zero
point was estimated to be about 0.05~magnitudes, which corresponds to
an uncertainty in the velocity field of $2.5\%$ of the distance.

To estimate the effects of such uncertainties we have run the
likelihood analysis with our tilted \lcdm test-case, conservatively
using zero-point changes of twice the estimated uncertainty,
$\epsilon=\pm 0.05$.  Figure~\ref{fig:0point} illustrates the effect
on the results for these cases.  The changes of zero-point appear to
shift the location of the maximum-likelihood values essentially along
the high-likelihood ridge.  The high ridge is not altered by much when
the zero point varies in this range.  The right panel shows the
resulting best-fit PS for the three different zero points, and lists
the corresponding values of $\sigma_8 \Omega^{0.6}$.  While the
variations in the zero-point systematically affect the PS, the changes
are not large; they fall within the range of the formal likelihood
errors, and are of the same order of the uncertainty associated with
the random distance errors (compare to Fig.~\ref{fig:tlcdmT60}).  It
is encouraging that the amplitude of the PS on intermediate scales ($k
\sim 0.1 \ihmpc$) is robust vis-a-vis changes in  in the zero point.
Similar results were obtained when using the other families of PS
prior models.

Similar to the uncertainty in the zero point of the distance
indicator, there is also an uncertainy associated with the slope of
the TF relation. This could lead to correlated errors in the inferred
distances and peculiar velocities, due to the fact that the average
linewidth of SFI galaxies slightly depends on distance (Wegner \etal
1999). However, the impact of this uncertainty on our results is even
slightly smaller than that of the uncertainty in the zero point,
perhaps because the SFI sample has been selected intentionally to
minimize the distance dependence of the line widths.

\subsection{Nonlinear Effects}
\label{subsec:nl}

A basic assumption in our analysis has been that linear gravitational
instability theory is adequate for the purpose of recovering the PS
from observed velocities on the scales of interest here. This is based
on the fact that in the mildly-nonlinear regime the velocity field is
approximated by linear theory better than the density field (basically
because the velocity is a spatial integral of the density and is
affected by fluctuations on larger scales). Indeed, the success of the
recovery of the PS from the mock catalogs leads us to believe that
this assumption is justified. However, one cannot rule out the possibility 
that some nonlinear effects are artificially reduced to some degree 
in the particle-mesh $N$-body simulation, and it is 
possible that the smooth shape of the linear PS as predicted for the
CDM family of models may fail to properly match the nonlinear features
that may be present on small scales in the real data.
Therefore, we discuss in this section possible nonlinear effects, 
which could manifest themselves in different forms. For example, as coherent 
motions associated with the non-linear evolution of the PS 
(\S~\ref{subsubsec:nlnl}), or as incoherent random motions, 
perhaps due to shell-crossing, which may be modeled as an additional 
velocity component of dispersion $\sigma_v$ (\S~\ref{subsubsec:sigmav}).

\subsubsection{Nonlinear Power Spectra}
\label{subsubsec:nlnl}

A way to include more properly nonlinear effects in our analysis is by
developing an approximation for the nonlinear evolution of the PS and
then incorporating it in the likelihood analysis. Such approximations
exist for the density power spectrum, $P_\delta$ (\eg, Peacock \&
Dodds 1994; Jain, Mo \& White 1995; Peacock \& Dodds 1996, hereafter
PD), but we need a similar approximation for the evolution of the
{\it velocity} power spectrum, $P_v$, which is the quantity we
actually confront with the data. A development and application
of such an approximation is beyond the scope of the present paper and 
will be presented later (Zehavi \etal 1999). Here, we summarize some 
relevant issues and illustrate the magnitude of such effects.
 
\begin{figure}[tbp]
{\includegraphics{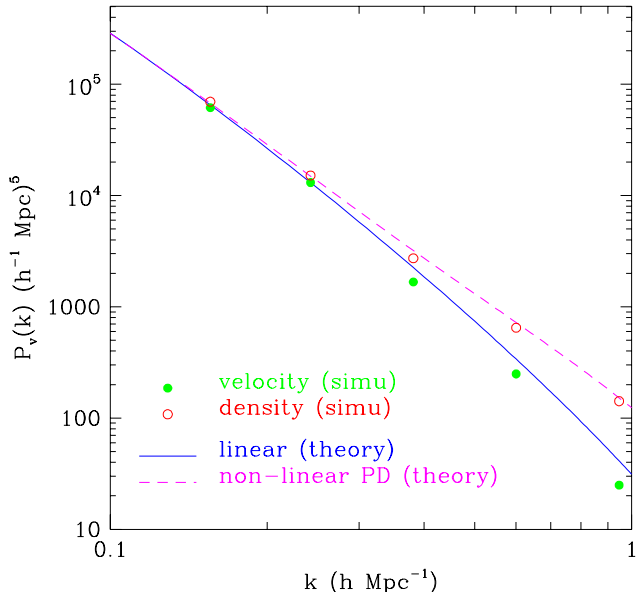}}
\vspace{7.6 cm}
\caption{\capt
The velocity PS as computed from the high-resolution simulation of
Colberg \etal (1999) (solid dots), compared with the theoretical
linear PS (solid line), the corrected PS using the PD formalism
(dashed line) and a computation via the density PS of the simulation
(open dots). The latter three have been transformed to velocity PS
using the linear velocity--density relation.  }
\label{fig:pvk}
\end{figure}

Figure~\ref{fig:pvk} shows the velocity PS computed in several 
different ways from an adaptive P$^3$M cosmological $N$-body simulation 
with a resolution higher by an order of magnitude than the simulation used 
in the present paper for the mock catalogs (but inside a smaller box of size 
$85 \hmpc$; GIF simulation, Colberg \etal 1999). The initial model used 
for $P_\delta$ is the so-called $\tau$CDM model with $\Omega=1.0$, $h=0.5$ 
and a modified shape parameter $\Gamma=0.21$. The figure clearly demonstrates 
that the velocity PS is reproduced by linear theory much better than the 
density PS. The $P_v$ that is computed directly from the evolved velocity 
field of the simulation (solid dots) lies slightly below the $P_v$ obtained 
from the assumed $P_\delta$ using linear theory ($P_v \propto k^{-2} 
P_\delta$, solid line).  On the other hand, the nonlinear correction to 
$P_\delta$ (\eg, PD) is larger than that of $P_v$ and in the opposite 
direction (upwards, as can be seen by the open dots and dashed line in 
Fig.~\ref{fig:pvk}).

One might have naively expected that the likelihood analysis using a 
pure linear treatment would be inferior to incorporating a non-linear
correction for $P_\delta$ followed by a linear translation to $P_v$.
However, as illustrated in Figure~\ref{fig:pvk}, this is not the case.
The latter procedure overestimates the nonlinear effects on the
velocity PS and increases the bias in the results. A similar bias is
reproduced when using the mock catalogs from the low-resolution
simulation of Kolatt \etal (1996), which exhibits a similar behavior
as in Fig.~\ref{fig:pvk}.  This could be remedied, in principle, by
incorporating the evolved $P_\delta$ and then counter-balancing it
with a proper approximation for the nonlinear velocity--density
relation, but this would be risky as we would be applying two large
corrections in opposite directions to mimic a small net effect. Until
we develop a direct nonlinear correction for $P_v$, we adopt the fully
linear procedure as our best approximation. This is justified by its
success in the mock catalogs and by the expectation for only small
nonlinear effects in $P_v$.

\subsubsection{Random Motions}
\label{subsubsec:sigmav}

We have made an ad-hoc attempt to model non-linearities by introducing
an uncorrelated velocity component of constant dispersion $\sigma_v$,
that adds a free term at zero-lag to the correlation function derived
from the linear PS model.  This may be a crude way to represent
small-scale random motions that are associated with
multi-streaming. An alternative interpretation of this additional
parameter may be as an unrecognized uncertainty in the distance
estimate which does not depend on distance and is therefore not
included in our usual error model. In either case, this provides a
test for the robustness of our results to an additional degree of
freedom.

\begin{figure}[tbp]
{\includegraphics{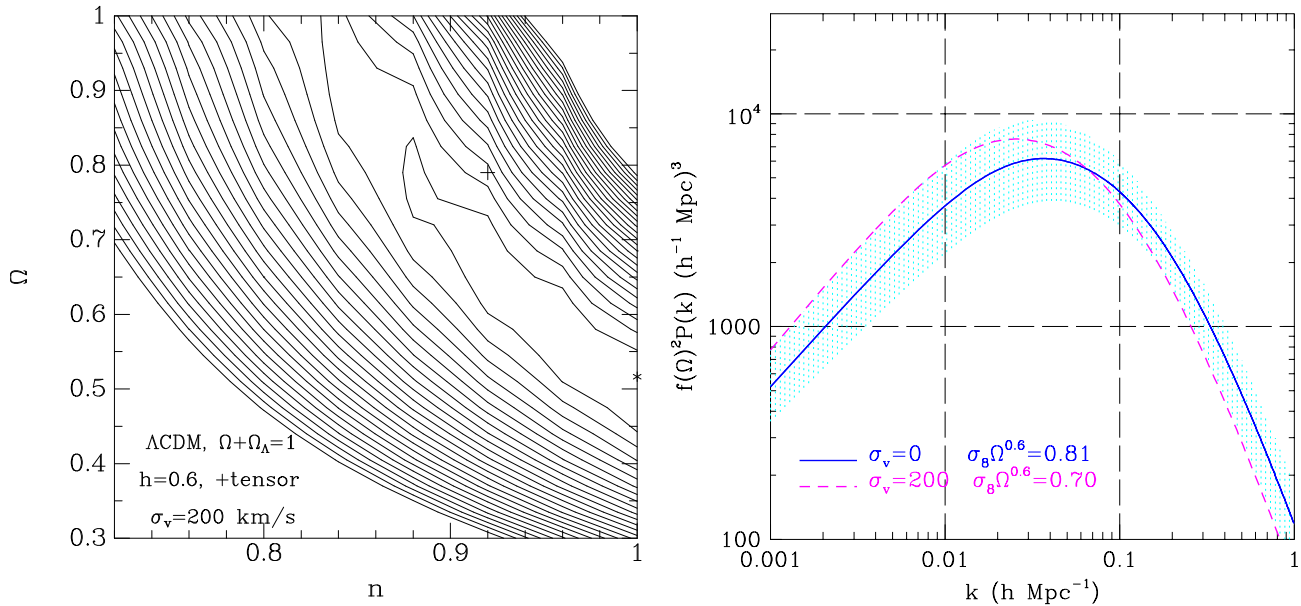}}
\vspace{7.7 cm}
\caption{\capt
Left panel: Likelihood contour plot in the $\Omega-n$ plane for the
tilted \lcdm $h=0.6$ model with tensor fluctuations, with an
additional scatter of $\sigma_v=200 \kms$. The maximum-likelihood
point is marked by `$*$', and the corresponding point for $\sigma_v=0$
is marked by `+'.  Right panel: The most likely PS when including
$\sigma_v=200$ in the fit (dashed line) and for the original
$\sigma_v=0$ case (solid line) together with its $90\%$ confidence
region shaded.  }
\label{fig:sigma_v}
\end{figure}

Figure~\ref{fig:sigma_v} demonstrates the effect of including a free
$\sigma_v$ in the likelihood analysis, again for our tilted \lcdm
test-case.  When allowing for this extra freedom, the preferred value
turns out to be $\sigma_v=200 \pm 120 \kms$, and is associated with a
PS that is slightly lower for $k>0.1$ and somewhat higher at small
$k$.  The value of $\sigma_8 \Omega^{0.6}$ is reduced by $14\%$.  The
deviations, in general, are comparable to the formal likelihood $90\%$
uncertainty marked by the shaded area. The likelihood contours are
somewhat sparser in this case, because of the additional scatter that
reduces the sensitivity to variations in the parameters. The ridge of
high likelihood is slightly shifted toward smaller values of the
cosmological parameters and it can now be roughly described by
\begin{eqnarray} 
\label{eq:tlcdmT60_sigv}
\qquad \Omega\, n^{3.9} ({\h60}^{1.3})& = &0.49\pm 0.09\,,
\end{eqnarray}  
a $\sim 15\%$ decrease compared to equation (\ref{eq:tlcdmT60}), which
is of the order of the random error.

Such a preference for a non-zero $\sigma_v$ associated with a change
in the PS is not recovered in the mock SFI catalogs, for which a
similar likelihood analysis turns out to prefer a negligible
$\sigma_v$ and a negligible effect on the PS. $N$-body simulations of
higher resolution may clarify this situation.

We note that the inclusion of a free $\sigma_v$ in the fit to the real
SFI data leads to results similar to those obtained when including a
free multiplicative parameter in the error model
(\S~\ref{subsec:tlcdmT60}).  The interpretation of a nonzero
$\sigma_v$ is thus not unique: It may refer to nonlinear effects that
exist in the real data but not in the current simulation, or it may
indicate that the actual errors are slightly larger than the original
estimates.  Since there is no clear benefit from adding this extra
parameter and the theoretical justification as a model for nonlinear
effects is weak, its inclusion in our main-stream analysis does not
seem to be justified. Still, in our total error-balance, we consider a 
systematic error of $15 \%$ due to non-linear effects.

\subsection{Comparison to the PS from Mark III}
\label{subsec:mark3}

A similar likelihood analysis (though with errors fixed a priori) 
has been recently applied by Zaroubi \etal (1997) to the Mark~III
catalog of peculiar velocities.  (Willick \etal 1995; 1996; 1997a).
It is interesting to investigate whether the recovered power spectra
are consistent with each other, given the respective uncertainties. 
This is intriguing because there are certain differences in the velocity 
fields as reconstructed from the two samples, especially in the bulk flows 
both in the very local neighborhood and of outer shells (\eg, da Costa \etal
1996; 1998; Dekel 1998; Dekel \etal 1998; Giovanelli \etal 1998a, 1998b).

\begin{figure}[tbp]
{\includegraphics{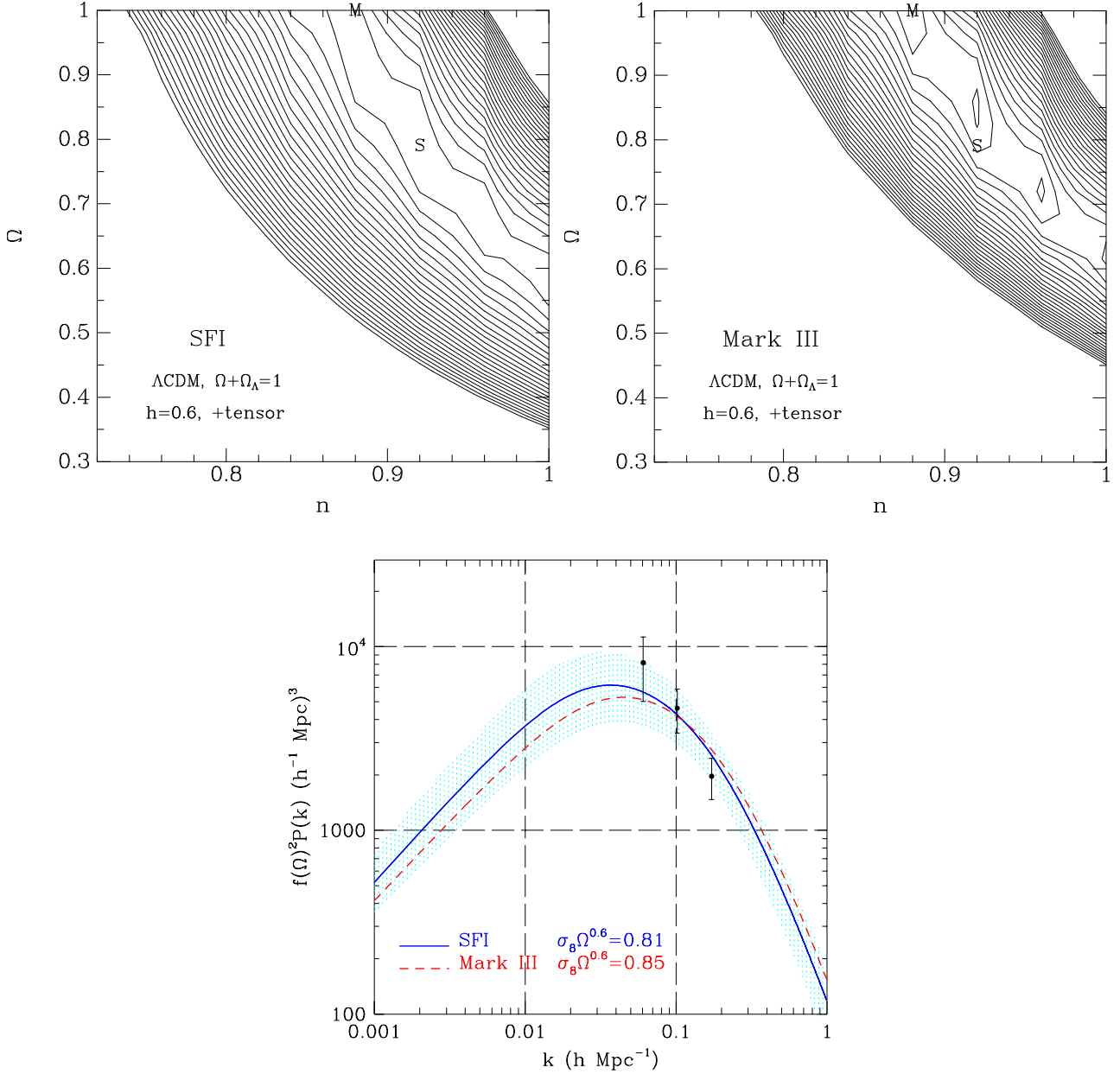}}
\vspace{14.5 cm}
\caption{\capt
Comparison of SFI PS results to Mark III for the tilted \lcdm
test-case. Top left panel shows $\ln\L$ contours for the SFI data, top
right panel shows the contours for the Mark III data. The best-fit
parameters for SFI and Mark III are marked, on both, by `S' and `M'
respectively. The lower panel shows the maximum likelihood PS
corresponding to SFI (solid) and Mark III (dashed). The 3 solid
dots mark the PS calculated from Mark III by Kolatt and Dekel (1997),
together with their estimated $1\sigma$ error-bar. The shaded region
is the SFI likelihood $90\%$ confidence region.  }
\label{fig:mark3}
\end{figure}

Figure~\ref{fig:mark3} presents a comparison of the likelihood
analysis for the two samples using our representative tilted \lcdm
family of models.  Shown once again are the likelihood contours for
the SFI data, together with the corresponding plot for the Mark III
data.  The high-likelihood ridge is similar for both samples. While
the Mark~III result slightly favors higher values of $\Omega$ and
lower values of $n$, the differences are along the ridge of maximum
likelihood and are therefore hardly significant.  The contours are
slightly more concentrated for the Mark III catalog because it
consists of more galaxies.

The best-fit PS for the two catalogs are shown in the bottom panel of
Figure~\ref{fig:mark3}, on top of the shaded area which marks the $90\%$
confidence region for SFI. The resultant power spectra are consistent
within the errors, and they agree particularly well on intermediate
scales, where the data provides the most meaningful constraints.  The
corresponding best values for $\sigma_8 \Omega^{0.6}$ are 0.81 and
0.85 for SFI and Mark~III respectively. Similar results are obtained
when comparing likelihood analysis of the two catalogs using the other
PS models.  It is worth noting here that the systematics discussed in
the previous sections, with regard to the SFI analysis, are found to
affect the Mark~III likelihood analysis in a similar way.  The PS
computed by Kolatt \& Dekel (1997) from the Mark~III smoothed density
field recovered by POTENT is also displayed on the figure (as three
symbols with error bars). The agreement of the SFI result with this
independent calculation of the Mark~III PS is good.

A recent comparison of Mark III with IRAS 1.2 Jy (Willick \& Strauss 1998)
suggests an alternative zero-point calibration for one of the Mark III
datasets. We have applied our likelihood analysis to the Mark III data
revised accordingly, and found negligible changes in the resulting power 
spectrum and cosmological parameters, smaller than the uncertainties due 
to global zero-point discussed in \S~\ref{subsec:0p}. 

The close agreement between the mass power spectra derived from the
two datasets indicates that the results presented here are quite
robust and are unlikely to arise form specific peculiarities of either
of the two samples.  This does not preclude possible differences that
are not picked up by the specific statistic used -- in our case, the
mass PS.  In particular, the difference in the two bulk flows, which
is known to exist, is not reflected in the power spectra.  This is
because the wavenumbers corresponding to the bulk velocity are smaller
than the $k$ range that dominates the fit in our current analysis.

\section{CONCLUSION}
\label{sec:concl}

We used a linear maximum-likelihood method to measure the mass-density
power spectrum from the SFI catalog of peculiar velocities, and to
determine the cosmological parameters for families of physical CDM
models with or without COBE normalization. We have corrected for
biases introduced by the non-trivial selection procedure of the SFI
catalog using a new semi-analytic procedure.  We have verified that
the results are quite insensitive to the detailed way by which we
implement this bias correction. This approach allows also to refine
the distance errors estimates. Our new version of likelihood analysis
enabled us to independently verify the error estimates of the SFI
catalog to within an uncertainty as small as $5\%$ of the error, which
we regard as very encouraging. Since the errors affect the PS in a
systematic way, this independent confirmation adds significantly to
our confidence in the results.

The general result for all the models examined here is that the power
spectrum at $k= 0.1 \ihmpc$ is $P(k) \Omega^{1.2} = (4.4\pm1.7)\times
10^3 \3hmpc$, and that $\sigma_8 \Omega^{0.6} = 0.82 \pm 0.12$.  These
results are obtained by the peculiar-velocity data independent of the
specific shape assumed for the PS, and are consistent with the result
of the $\Gamma$ model independent of the COBE normalization. The
random errors quoted are $90\%$ confidence level and they include
small variations due to the choice of model for the PS within the
families of models tried here.

For the general family of COBE-normalized CDM models, we find a
high-likelihood ridge in the $\Omega-n-h$ parameter space, which can
be crudely approximated by $\Omega\, n^{\nu}\, {\h60}^{\mu} = 0.62 \pm
0.15$, where for \lcdm $\mu=1.3$ and $\nu=2.0,3.9$, without and with
tensor fluctuation respectively. For \ocdm, without tensor
fluctuations, the powers are $\mu=0.9$ and $\nu=1.4$. Again, the error
quoted is the formal $90\%$ uncertainty including the model
variations.  Thus, for $h=0.6$, the maximum-likelihood value of
$\Omega$ ranges between 0.6 and unity while $n$ varies between 1 and
0.8 respectively.  Without a tilt, values of $\Omega$ as low as 0.5
are allowed within the $90\%$ confidence limit.

Our tests using mock catalogs based on an $N$-body simulation that
mimics our cosmological neighborhood indicate that the systematic
errors in our results are relatively small. In particular, the
nonlinear effects in the mock catalogs are found to be
negligible. This is indeed expected because the quantity we actually
measure is the velocity power spectrum in the mildly-nonlinear regime,
which we have demonstrated to be reasonably approximated by linear
theory. An ad-hoc test for nonlinear (multi-streaming) effects in the
data themselves indicated that they may work to reduce the values of
the cosmological parameters given above, but that this effect is not
larger than $\sim 15\%$. In order to refine our estimates of the
systematic effects even further, we intend to repeat the current
analysis using a proper nonlinear scheme, and to repeat the tests of
the method using simulations of higher resolution which are in
preparation.  We thus estimate the total systematic uncertainty to be
of order $\sim 15\%$, namely comparable in size to the random
errors. Therefore, to be on the safe side when comparing our results
to other results, we recommend as a rule of thumb multiplying the
quoted errors by a factor of $\sim 2$.

As yet another word of caution, it is worth recalling that our
analysis is heavily weighted by the galaxies at relatively small
distances, because the data is weighted by the inverse squared of the
distance errors. This means that the result is sensitive to the data
and error estimate of the inner galaxies. It is possible in principle
that a source of distance error which operates preferentially at small
distances has somehow escaped our attention and is not properly
modeled by our error model.  To test the effect of such a possibility,
we have repeated the analysis after pruning all galaxies with
distances smaller than a given distance.  When pruning inside
$15\hmpc$ ($3\%$ of the data) we obtain for the most likely value
$\sigma_8\Omega^{0.6} =0.85$ instead of the original result of
$\sigma_8\Omega^{0.6} =0.81$ when using all the data (still with our
standard pruning based on linewidth).  When pruning inside $25\hmpc$
($17\%$ of the galaxies), we obtain instead $\sigma_8\Omega^{0.6}
=0.71$.  It is encouraging to find that these variations are within
the $90\%$ likelihood contours of the different cases in the $\Omega-n$
plane, but this is yet another potential source of uncertainty to bear
in mind.

A systematic trend does seem to show up when we eliminate as much as
the whole inner half of the data (inside a distance of $46\hmpc$, or
with linewidth smaller than $2.48$); the outer data, when analyzed by
themselves, indicate a significantly lower PS then the inner data.
This effect is not reproduced in the mock catalogs and is therefore
not likely to represent a general fault in the method.  Possible
explanations for this effect are larger uncertainty in our estimate of
random and systematic errors at large distances, differences between
the assumed and the true TF relation, or a true PS with a different
shape than our models.  It may also be due to a real difference
between the density fields in the two halves (that is somehow not
properly reproduced in the simulation) or to a systematic dependence
of velocity bias on galaxy properties.  We carried out a number of
tests in which we added to the likelihood analysis ad-hoc free
parameters which allow more flexibility in the distance dependence.
These include variations in the TF parameters and in the errors as a
function of linewidth. Our tests indeed led to some improvement in the
agreement between the two subsamples, but, being only preliminary,
they did not yield so far a firm conclusion as for the dominant source
of the effect and the optimal way to deal with it.  Since the
variations introduced preferentially affect the peculiar velocities of
galaxies at large distances, which typically have large errors and
therefore contribute only little to the likelihood procedure, they do
not affect significantly the resultant power spectrum from the full
sample. We therefore conclude that our current results are robust, and
defer a more thorough investigation of this trend to a future
analysis.

The recovered mass power spectrum, and the constraints on the
cosmological parameters obtained here, are consistent with the results
of a similar analysis applied to the Mark~III catalog of peculiar
velocity. This is despite the fact that these two catalogs seem to
differ in some of their other properties, such as the large-scale bulk
velocity.  Indeed, the bulk velocity is not expected to contribute to
the density on smaller scales.  There is also an apparent disagreement
between the results obtained from peculiar velocities of clusters
(Borgani \etal 1997) and our result for the SFI field galaxies.

As mentioned in the Introduction, our dynamical result of
$\tilde\sigma_8 \equiv \sigma_8 \Omega^{0.6} \simeq 0.8\pm 0.2$ may be
crudely compared to estimates of the $\beta$ parameter obtained when
comparing the same SFI data to a redshift survey of galaxies. da Costa
\etal (1998) find $\beta=0.6\pm 0.1$ when comparing the SFI peculiar
velocities to the velocities predicted by the IRAS 1.2 Jy redshift
survey, assuming linear biasing.  A similar value was obtained from
Mark III when the comparison was done via velocities (Davis \etal 1996).
With $\sigma_{8\rm g}
\simeq 0.7$ for IRAS galaxies, the predicted $\beta$ from our current
constraint on $\tilde\sigma_8$, via $\beta=\tilde\sigma_8 /
\sigma_{8\rm g}$, is significantly closer to unity 
than to 0.6 (compare also to
Kolatt \& Dekel 1997, Fig. 6).  The residuals between the measured
peculiar velocities and the IRAS predictions, for the best-fit $\beta$
value, were found in this comparison (da Costa \etal 1998) to be
significantly higher than the errors as originally estimated for the 
SFI data (based on the scatter observed in the SCI cluster sample)
combined with the errors estimated for the IRAS data.

One possibility is that the the IRAS model fails to predict some of
the peculiar velocities that exist in the SFI data, e.g., because the
distribution of galaxies is not properly approximated by a simple,
linear, scale-independent and deterministic biasing relation (\eg,
Dekel \& Lahav 1998). In that case, the interpretation of the value of
$\beta$ determined from fitting the IRAS predictions to the velocity
data is not clear.
 
On the other hand, it is also possible that the errors in SFI are
indeed larger than originally estimated. Such larger errors would
accordingly reduce the PS amplitude, in particular the value of
$\tilde\sigma_8$, as estimated in the current paper. However, such an
effect should have been detected by investigating the global biasing
properties of the sample. It would also be hard to understand why our
likelihood analysis does prefer errors very similar to the original
estimates. Although we are fairly convinced that nonlinear effects in
the current analysis are confined to the level of $\leq 15\%$, it will
be worth making an extra effort to improve the accuracy in a future
paper.

\acknowledgments{We thank George Blumenthal, Stefano Borgani, Gerard Lemson,
Adi Nusser, and Michael Strauss for stimulating discussions.  We are
grateful to the ESO Visitors fund for supporting visits to Garching by
AD, GW, IZ, MPH, and RG. This research was supported in part by an ESO
DGDF grant to WF, the US-Israel Binational Science Foundation grant
95-00330 and the Israel Science Foundation grant 950/95 to AD, IZ and
AE, the DOE and the NASA grant NAG 5-7092 at Fermilab to IZ, NSF
grants AST94-20505 and AST96-17069 to RG, AST95-28860 to MPH, and
AST93-47714 to GW.}

{}


\begin{thebibliography}{}

\bibitem{bbks} Bardeen, J. M., Bond, J. R., Kaiser, N., \& Szalay, A. S. 1986,
\apj, 304, 15
  
\bibitem{cobe4yr} Bennett, C. L., \etal 1996, \apj, 464, L1

\bibitem{bo} Bothun, G. D., Mould, J. R. 1987, \apj, 313, 629

\bibitem{sci_vrms} Borgani, S., da Costa, L. N., Freudling, W., Giovanelli, 
R., Haynes, M. P., Salzer, J. \& Wegner, G., 1997, \apj, 482, L121

\bibitem{gif} Colberg, J. M. \etal 1999, in preparation

\bibitem{sfi_pot} da Costa, L. N., Freudling, W., Wegner, G., Giovanelli, R., 
Haynes, M. P., \& Salzer, J. J. 1996, \apj, 468, L5

\bibitem{sfi_itf} da Costa, L. N., Nusser, A., Freudling, W., Giovanelli, R., 
Haynes, M. P., Salzer, J. J., \& Wegner, G. 1998, \mnras, 299, 425

\bibitem{m3_itf} Davis, M., Nusser, A., \& Willick, J. A. 1996, \apj, 473, 22

\bibitem{dekel} Dekel, A. 1998, in {\it Formation of Structure in the
Universe}, eds. A.  Dekel \&  J. P. Ostriker, Cambridge: Cambridge University
Press, 250

\bibitem{pot} Dekel, A., Bertchinger, E., \&  Faber, S. M. 1990, \apj, 364, 349

\bibitem{m3_pot} Dekel, A., Eldar, A.,  Kolatt, T., Yahil, A., Willick, J. A.,
Faber, S. M., Courteau, S., \& Burstein, D. 1998, 
\apj, submitted (astro-ph/9812197) 

\bibitem{biasing} Dekel, A., \&  Lahav, O.  1998, \apj, submitted 
(astro-ph/9806193) 

\bibitem{gamma} Efstathiou, G., Bond, J. R., \& White, S. D. M. 1992, \mnras, 
258, 1p

\bibitem{eldar} Eldar, A. \etal 1999, in preparation

\bibitem{iras_1.2} Fisher, K. B., Huchra, J. P., Strauss, M. A., Davis, M., 
Yahil, A., \& Schlegel, D. 1995, \apjs, 100, 69

\bibitem{oi} Freudling, W., da Costa, L. N., Pellegrini, P. S. 1994, \mnras,
268, 943

\bibitem{mc_mb} Freudling, W., da Costa, L. N., Wegner, G., Giovanelli, R., 
Haynes, M. P., \& Salzer, J. J. 1995, \aj, 110, 920

\bibitem{herc} Freudling, W., Martel, H., Haynes, M. P.  1991, \apj, 377, 349

\bibitem{sci} Giovanelli, R., Haynes, M. P., Herter, T., Vogt, N., Salzer, J. 
J., Wegner, G., da Costa, L. N., \& Freudling, W. 1997a, \aj, 113, 22

\bibitem{sci2} Giovanelli, R., Haynes, M. P., Herter, T., Vogt, N., da Costa, 
L. N., \& Freudling, W., Salzer, J. J., Wegner, G. 1997b, \aj, 113, 53 

\bibitem{dip2} Giovanelli, R., Haynes, M. P., Freudling, W., Da Costa, L. N., 
Salzer, J. J., \& Wegner, G. 1998a, \apj,  505, L91
 
\bibitem{dip1} Giovanelli, R., Haynes, M. P.,  Salzer, J. J.,  Wegner, G., 
Da Costa, L. N., \& Freudling, W. 1998b, \aj, 116, 2632

\bibitem{gor} G\'orski, K. M. 1988, \apj, 332, L7

\bibitem{psi} G\'orski, K. M., Davis, M., Strauss, M. A., White, S. D. M., 
\& Yahil, A. 1989, \apj, 344, 1  

\bibitem{cobe3} G\'orski, K. M., Ratra, B., Stompor, R., Sugiyama, N., 
\& Banday, A. J. 1998, \apjs, 114, 1

\bibitem{cobe1} G\'orski, K. M., Ratra, B., Sugiyama, N., \& Banday, A. J. 
1995, \apj, 444, L65

\bibitem{gro} Groth, E. J., Juszkiewicz, R., \& Ostriker, J. P. 1989, \apj, 
346, 558


\bibitem{cobe} Hinshaw, G., Banday, A. J., Bennett, C. L., Gorski, K. M.,
Kogut, A., Smoot, G. F., \& Wright, E. L. 1996, \apjl, 464, L17

\bibitem{sfi_dat1} Haynes, M. P. \etal 1999, in preparation

\bibitem{jk} Jaffe, A. H., \& Kaiser, N. 1995, \apj, 455, 26 

\bibitem{jmw_nl} Jain, B., Mo., H. J., \& White,  S. D. M. 1995, \mnras, 276, 
L25

\bibitem{kash} Kashlinsky, A. 1998, \apj, 492, 1

\bibitem{kaiser} Kaiser, N. 1988, \mnras, 231, 149

\bibitem{kof} Kofman, L., Bertschinger, E., Gelb, J. M., Nusser, A., \&
Dekel, A. 1994, \apj, 420, 44

\bibitem{kd} Kolatt, T., \& Dekel, A. 1997, \apj, 479, 592

\bibitem{mocks} Kolatt, T., Dekel, A., Ganon, G., \& Willick, J. A. 1996, 
\apj, 458, 419

\bibitem{math} Mathewson, D. S., Ford, V. L., \& Buchhorn, M. 1992, \apjs, 81,
413

\bibitem{pd94} Peacock, J. A., \& Dodds, S. J. 1994, \mnras, 267, 1020

\bibitem{pd96} Peacock, J. A., \& Dodds, S. J. 1996, \mnras, 280, L19

\bibitem{potiras} Sigad,Y., Dekel, A., Eldar, A., Strauss, M., Yahil, A. 1998,
\apj, 495, 516 

\bibitem{strauss} Strauss, M. A. 1998, in {\it Formation of Structure in the 
Universe}, eds. A.  Dekel \&  J. P. Ostriker, Cambridge: Cambridge University 
Press, 172

\bibitem{sw} Strauss, M. A., \& Willick, J. A. 1995, Phys. Rep., 261, 271

\bibitem{sugi} Sugiyama, N. 1995, \apjs, 100, 281

\bibitem{omega_b} Tytler, D, Fan, X-M, \& Burles, S. 1996, Nature, 381, 207

\bibitem{sfi_dat2} Wegner, M. P. \etal 1999, in preparation

\bibitem{cobe2} White, M., \& Bunn, E. F. 1995, \apj, 450, 477

\bibitem{w1} Willick, J. A., Courteau, S., Faber, S. M., Burstein, D., \&
Dekel, A. 1995, \apj, 446, 12

\bibitem{w2} Willick, J. A., Courteau, S., Faber, S. M., Burstein, D.,
Dekel, A., \& Kolatt, T. 1996, \apj, 457, 460

\bibitem{w3} Willick, J. A., Courteau, S., Faber, S. M., Burstein, D.,
Dekel, A., \& Strauss, M. A. 1997a, \apjs, 109, 333

\bibitem{velmod2} Willick, J. A., \&  Strauss, M. A. 1998, \apj, 
507, 64 

\bibitem{velmod} Willick, J. A., Strauss, M. A., Dekel, A., \& Kolatt, T. 
1997b, \apj, 486, 629

\bibitem{z} Zaroubi, S., Zehavi, I., Dekel, A., Hoffman, Y., \& Kolatt, T. 
1997, \apj, 486, 21

\bibitem{vel_nl} Zehavi, I. \etal 1999, in preparation

\end{thebibliography}
\end{document}